\batchmode
\makeatletter
\def\input@path{{\string"/Users/axel/Desktop/paper cev/paraqlosubasaarxivyparatirarchivos___concoverlett/\string"}}
\makeatother
\documentclass[english]{extarticle}
\usepackage{lmodern}
\usepackage[T1]{fontenc}
\usepackage[latin9]{inputenc}
\usepackage{geometry}
\usepackage{array}
\usepackage{float}
\usepackage{units}
\usepackage{textcomp}
\usepackage{multirow}
\usepackage{amstext}
\usepackage{amssymb}
\usepackage{graphicx}
\usepackage{esint}

\makeatletter

\providecommand{\tabularnewline}{\\}

\usepackage{hyperref}
\usepackage[numbers,sort&compress]{natbib}
\date{}

\@ifundefined{showcaptionsetup}{}{%
 \PassOptionsToPackage{caption=false}{subfig}}
\usepackage{subfig}
\makeatother

\usepackage{babel}
\begin{document}

\title{\textbf{Computing the CEV option pricing formula using }\\
\textbf{the semiclassical approximation of path integral}}

\author{Axel A.~Araneda\thanks{Faculty of Engineering \& Sciences, Universidad Adolfo Ib\'a\~nez,
Avda. Diagonal las Torres 2640, Peñalol\'en, 7941169 Santiago, Chile.
Email: axel.araneda@edu.uai.cl.}$\quad$and Marcelo J.~Villena\thanks{Universidad Adolfo Ibáñez, Address: Diagonal Las Torres 2640, Peñalolén,
Santiago, Chile. Telephone: 56-223311491, Email: marcelo.villena@uai.cl.} \thanks{Aid from the Fondecyt Program, project Nº 1131096, is grateful acknowledged
by Marcelo. }}
\maketitle
\begin{center}
\vspace{-1em} Last version: \today \vspace{2em}
\par\end{center}
\begin{abstract}
The Constant Elasticity of Variance (CEV) model significantly outperforms
the Black-Scholes (BS) model in forecasting both prices and options.
Furthermore, the CEV model has a marked advantage in capturing basic
empirical regularities such as: heteroscedasticity, the leverage effect,
and the volatility smile. In fact, the performance of the CEV model
is comparable to most stochastic volatility models, but it is considerable
easier to implement and calibrate. Nevertheless, the standard CEV
model solution, using the non-central chi-square approach, still presents
high computational times, specially when: i) the maturity is small,
ii) the volatility is low, or iii) the elasticity of the variance
tends to zero. In this paper, a new numerical method for computing
the CEV model is developed. This new approach is based on the semiclassical
approximation of Feynman's path integral. Our simulations show that
the method is efficient and accurate compared to the standard CEV
solution considering the pricing of European call options. 
\end{abstract}
\begin{center}
Keywords: Option pricing, constant elasticity of variance model, path
integral, numerical methods. 
\par\end{center}

\vfill{}
\pagebreak{}

\section{Introduction}

One of the most significant limitations of the Black-Scholes (BS)
\cite{black1973pricing} model is the assumption of constant volatility,
which ignores some well-known empirical regularities such as: the
leverage effect \cite{bekaert2000asymmetric,bollerslev2006leverage},
and the volatility smile \cite{rubinstein1985nonparametric,jackwerth1996recovering}.
These shortcomings have inspired several non-constant volatility models
in continuous time\footnote{For discrete-time approaches to modeling volatility, see refs. \cite{engle1982autoregressive,bollerslev1992arch} },
considering `stochastic volatility'\footnote{A comprehensive review for stochastic volatility models can be found
in \cite{Ghysels1996119} and \cite{shephard2009stochastic} .} or `level-dependent volatility' models\footnote{A.k.a. `local volatility'}
\cite{hobson1998complete}. In the former, both the asset and the
volatility have their own diffusion processes. In the level-dependent
volatility models only the asset is governed by a diffusion process,
and its volatility is modeled in function of the asset level. In this
paper, the analysis will be focused on the the constant elasticity
of variance (CEV) model, proposed by J. Cox, the most known level-dependent
volatility approach \cite{Cox1975,cox1996constant}.

Furthermore, the CEV model has a marked advantage in capturing basic
empirical regularities such as: heteroscedasticity, the leverage effect,
and the volatility smile\cite{beckers1980constant,schroder1989computing,davydov2001pricing,chen2015equity}.
As a consequence, the CEV model significantly outperforms the Black-Scholes
(BS) model in forecasting both prices and options \cite{macbeth1980tests,emanuel1982further,tucker1988tests,lauterbach1990pricing,singh2011forecasting}.
Furthermore, the performance of the CEV model is comparable to most
stochastic volatility models, but it is considerable easier to implement
and calibrate \cite{yuen2001estimation}. 

In terms of the option pricing using the CEV model, the exact formula
for a vanilla European option involves a complex computation of an
infinite series of incomplete gamma functions \cite{Cox1975}. Subsequently,
\cite{schroder1989computing} matched the Cox pricing formula with
the non-central chi-square distribution. Schroder also provides a
simple approximated method for its computation, see \cite{hsu2008constant}
for a detailed derivation of the two methodologies. Since then, the
use of the non-central chi-square distribution becomes the most widely
used method of pricing for options under the CEV model. Besides, several
alternative methods for its implementation have been developed \cite{larguinho2013computation}. 

Nevertheless, the standard CEV model solutions, using the non-central
chi-square approach, still presents considerably high computational
times, specially when: i) the maturity is small, ii) the volatility
is low, or iii) the elasticity of the variance tends to zero\cite{schroder1989computing,thakoor2013new,thakoor2015fast}.
In order to deal with these problems, many approaches have been reported
for the European-vanilla type option pricing. These approaches include
numerical schemes \cite{lu2005valuation,thakoor2013new,aboulaich2014stochastic,cruz2017binomial},
Montecarlo simulations \cite{lindsay2012simulation}, perturbation
theory model \cite{park2011asymptotic}, and analytical approximations
to the transition density \cite{pagliarani2012analytical} or to the
hedging strategy \cite{krasin2017approximate}, among others.

In this paper, a new numerical method for computing the CEV model
is developed. This new approach is based on the semiclassical approximation
of Feynman's path integral model. In financial literature, path integral
techniques have already been used in the option pricing problem, see
\cite{linetsky1997path,montagna2002path,bormetti2006pricing,baaquie2007quantum,lemmens2008path,devreese2010path}.
Nevertheless, the main focus has been in theoretical issues rather
than in practical applications. On the other hand, the application
of the semiclassical approximation of Feynman's path integral technique
on financial problems is rather limited, see for example \cite{contreras2010quantum,contreras2014stochastic,kakushadze2015path}. 

\cite{kakushadze2015path} points out that, just as in the case of
quantum mechanics, the path integral approach in finance is neither
a panacea, nor is it intended to yield fundamentally new results,
but in some cases it provides clarity and insight into old problems.
In this paper, we analyze the possibility that the path integral approach
could also be an interesting computational tool to solve complex problems
in quantitative finance. In this context, this research could be important
not only because it develops a novel and efficient technique for the
solution of the renowned CEV model, but also because it could open
the door to computational applications of such methods of quantum
mechanics. Indeed, our simulations show that computing the CEV option
pricing formula using the semiclassical approximation of path integral
is efficient and accurate compared to the standard CEV solution considering
the pricing of European call options. Additionally, the proposed approximation
reduces execution times importantly and keep the simplicity of the
traditional solution. 

Thus, the main idea of the paper is twofold, firstly to use ideas
from quantum mechanics to deal with applied finance problems, and
secondly, to develop practical methodologies and to test them numerically
in specific case studies, while discussing their practical advantage
and limitations. The structure of the paper is the following. Firstly,
Feynman's path integral formulation is revisited. Secondly, the path
integral approximation is applied to the basic BS model. Thirdly,
the path integral approximation is applied to the CEV model. Later,
a numerical solution to the CEV model is developed. In the next section,
several numerical simulations are carried out in order to measure
the performance of the new method, comparing the path integral approximation
with the traditional non-central chi-squared approach for the pricing
of European call options. Finally, some conclusions and future research
avenues are outlined.

\section{The Feynman path integral approach}

The path integral formalism was developed by Richard P. Feynman \cite{feynman1948space},
introducing the action principle from classical mechanics to quantum
mechanics. Nowadays Feynman's path integral is a well-known tool in
quantum mechanics, and statistical and mathematical physics, with
applications in many branches of physics such as: optics, thermodynamics,
nuclear physics, atomic and molecular physics, cosmology, polymer
science and other interdisciplinary areas \cite{feynman2010quantum,1998handbook}. 

In the following lines, we describe the fundamentals of the path integral
methodology. The starting point is the Schr\"{o}dinger equation:

\begin{equation}
i\hbar\frac{\partial\Psi}{\partial\tilde{t}}=\mathrm{\hat{H}}_{BS}\Psi\label{eq: Schrodinguer}
\end{equation}

\noindent where $\Psi$ is the wave function and $\hat{\mathrm{H}}$
the Hamiltonian quantum operator (for this instance we consider a
time independent Hamiltonian).

Considering $\Psi_{0}(x)$ as the initial value of $\Psi$ (i.e.,
$\Psi(x,t=0)=\Psi_{0}(x)$), the general solution of \ref{eq: Schrodinguer BS}
is given in term of the unitary evolution operator:

\begin{eqnarray}
\Psi(x,t) & = & e^{\nicefrac{-i\hat{\text{H}}t}{\hbar}}\Psi_{0}(x)\label{eq:evolution}
\end{eqnarray}
\\
Equivalently, using convolution properties, the value at time $t$
of the wave function is represented by:

\begin{eqnarray}
\Psi(x,t) & = & \int_{-\infty}^{\infty}\text{e}^{-i\hat{\text{H}}t/\hbar}\delta\left(x-x_{0}\right)\Psi_{0}\left(x_{0}\right)\mathrm{d}x_{0}\nonumber \\
 & = & \int_{-\infty}^{\infty}K\left(x,t\,|\,x_{0},0\right)\Psi_{0}\left(x_{0}\right)\mathrm{d}x_{0}\label{eq: prop_Sch}
\end{eqnarray}
\\

\noindent where $K\left(x,t|x_{0},0\right)=<x_{0}|e^{-i\mathrm{t\hat{H}/\hbar}}|x>$
is called the propagator. 

Feynman concentrated on a previous work of Dirac \cite{Dirac1933},
related with the proportionality between the exponential of the action
over the classical path (which come from the Lagrangian formalism)
and the propagator in quantum mechanics:

\[
K\left(x,t\,|\,x_{0},0\right)\propto\text{e}^{(i/\hbar)A\left[x_{cl}\right]}
\]

\noindent where $A$ is the action functional, defined as the time
integral Lagrangian:

\[
A[x(t)]=\int_{0}^{t}\mathcal{L}(x,\dot{x},t')\text{d}t'
\]

\noindent $A\left[x_{cl}\right]$ indicates that the action is evaluated
over the classical trajectory from $x_{0}$ to $x$.

Feynman reformulated Dirac formulation and described the propagator
as the contributions of the all virtual paths, not only the classical
ones:

\[
K\left(x,t\,|\,x_{0},0\right)=\sum_{\begin{array}{c}
\text{All Paths}\\
\text{from \ensuremath{x_{0}} to \ensuremath{x}}
\end{array}}\tilde{\mathcal{N}}\text{\ensuremath{\text{e}^{(i/\hbar)A\left[x(t)\right]}}}
\]

\noindent where $\tilde{\mathcal{N}}$ is an appropriated normalization
for $K$. 

Thus, using the Riemann integral for each path (see ref. \cite{feynman2010quantum}),
the propagator is defined as:

\begin{equation}
K\left(x,t\,|\,x_{0},0\right)=\int\text{\ensuremath{\mathcal{D}}\ensuremath{\ensuremath{\left[x(t)\right]}}}\text{\ensuremath{\text{e}^{(i/\hbar)A\left[x(t)\right]}}}\label{eq:Path integral}
\end{equation}

The functional integral of the right-hand side of the Eq. \ref{eq:Path integral}
is defined as a `Path Integral', and the measure of the integration
is given by $\mathcal{D}\ensuremath{\left[x(t)\right]}$ which means
the integrations over all trajectories.

The computation of the path integral is done via the time slicing
scheme \cite{feynman1948space,feynman2010quantum}, which is not a
straightforward procedure. Nevertheless, there is an alternative and
popular method used in physics called the `semiclassical approximation',
which approximates the argument of the path integral into a Gaussian
function, arriving this way to a solution in terms of the classical
path, see \cite{rajaraman1975some,koeling1975semi,kakushadze2015path}.
Since the aim of the paper is to find a more efficient numerical solution
to a complex problem, this avenue seems plausible and attractive,
see \cite{kakushadze2015path} for a presentation of the semiclassical
approximation to option pricing. The general procedure is explained
below. 

First, we write the path that links the points $x(t_{0})=x_{0}$ with
$x(t_{1})=x_{1}$ as the classical trajectories as the main contribution
plus the fluctuations around it:

\begin{equation}
x(t)=x(t)_{cl}+\delta x(t)\label{eq: paths}
\end{equation}

\noindent with the fixed conditions (extremality condition):

\begin{equation}
\delta x(t_{0})=\delta x(t_{1})=0\label{eq:cond}
\end{equation}

Later, we can expand the action around to $x_{cl}(t)$ using a functional
Taylor series \cite{chaichian2001path}:

\begin{eqnarray}
A\left[x(t)_{cl}+\delta x\right] & = & A\left[x(t)\right]\biggr|_{x_{cl}(t)}+\int_{t_{0}}^{t_{1}}\text{d}t\frac{\delta A\left[x(t)\right]}{\delta x(t)}\biggr|_{x_{cl}(t)}\delta x(t)\label{eq:expansion}\\
 &  & +\frac{1}{2}\int_{t_{0}}^{t_{1}}\text{d}t\text{d}t'\frac{\delta^{2}A}{\delta x(t)\delta x(t')}\delta x(t)\delta x(t')\biggr|_{x_{cl}(t)}\\
 &  & +\frac{1}{3!}\int_{t_{0}}^{t_{1}}\text{d}t\text{d}t'\text{d}t''\frac{\delta^{2}A}{\delta x(t)\delta x(t')\delta x(t'')}\delta x(t)\delta x(t')\delta x(t'')\biggr|_{x_{cl}(t)}+\mathcal{O}(4)
\end{eqnarray}

The semiclassical approximation consist in truncated up to the quadratic
terms the expansion \ref{eq:expansion}:

\[
A\left[x(t)\right]\approx A\left[x_{cl}(t)\right]+\frac{1}{2}\int_{t_{0}}^{t_{1}}\frac{\delta^{2}A}{\delta x(t)\delta x(t')}\delta x(t)\delta x(t')\biggr|_{x_{cl}(t)}
\]

\noindent  where the linear term is vanished due to the extremality
condition. 

Thus, the propagator in the semiclassical limit becomes:

\begin{eqnarray}
K^{SC}\left(x_{1},t_{1}|x_{0},t_{0}\right) & = & \text{e}^{(i/\hbar)A\left[x_{cl}(t)\right]}\int_{x_{T}}^{x}\mathcal{D}\left[\chi(t)\right]\text{e}^{(i/\hbar)\frac{1}{2}\int_{t_{0}}^{t_{1}}\frac{\delta^{2}A}{\delta x(t)\delta x(t')}\delta x(t)\delta x(t')\biggr|_{x_{cl}(t)}}\nonumber \\
 & = & \text{e}^{(i/\hbar)A\left[x_{cl}(t)\right]}\mathcal{N}\label{eq:semi-1}
\end{eqnarray}
\\

\noindent where $\mathcal{N}$ is a normalization constant which
incorporates the contribution of the second order term, defined by
a Gaussian path integral . An analytical expression was developed
for it in ref. \cite{morette1951definition} as the necessary condition
to maintain the unitary measure of the probability amplitudes \cite{dewitt1974feynman},
and it's equal to:

\begin{equation}
\mathcal{N}=\sqrt{-\frac{\mathcal{M}}{2\pi}}\label{eq:normalization}
\end{equation}

\noindent where $\mathcal{\mathcal{M}}$ is the the van Vleck-Pauli-Morette
determinant\footnote{A.k.a Morette-Van Hove determinant. See ref. \cite{choquard1996story}
for details} \cite{morette1951definition,van1928correspondence}, computed as:

\begin{eqnarray}
\mathcal{M} & = & \frac{\partial^{2}A_{class}}{\partial y_{0}\partial y_{T}}\label{eq:vanVleck-1}
\end{eqnarray}

Finally, in the semiclassical regime, the propagator becomes\footnote{The Eq. \ref{eq:K semi-1} is called the Pauli formula \cite{1998handbook}}:

\begin{equation}
K\left(x,t\,|\,x_{0},0\right)=\sqrt{-\frac{\mathcal{M}}{2\pi}}\text{e}^{\frac{i}{\hbar}A[x_{cl}]}\label{eq:K semi-1}
\end{equation}

The only necessary condition to get a solution for Eq. \ref{eq:K semi-1}
is to have an analytical expression for the action over the classical
path. This can be achieved via the Hamilton equations (or Euler-Lagrange
equation) using the classical Hamiltonian related to the quantum Hamiltonian
defined in \ref{eq: Schrodinguer}.

Finally, two important notes must be considered in relation to the
semiclassical approximation \cite{1998handbook}:
\begin{enumerate}
\item[i)] It is exact if the Lagrangian is quadratic\label{quadratic}.
\item[ii)] It satisfy the Schr\"{o}dinger equation up to terms of order $\hbar^{2}$.
\end{enumerate}
In the next section, we apply the semiclassical approximation of path
integral to the European-vanilla type option pricing, arriving to
the famous Black-Scholes model.

\section{A semiclassical approximation of the path integral approach to the
Black-Scholes model\label{sec:The-Black-Scholes}}

We assume stochastic spot prices $S_{t}$, governing by a standard
geometric Brownian motion under the physical $\mathbb{P}$-measure
of the form:

\begin{equation}
\frac{\mathrm{d}S_{t}}{S_{t}}=u\mathrm{d}t+\sigma\mathrm{d}\tilde{W}_{t}\label{eq:GBM}
\end{equation}

\noindent where $W_{t}$ is a standard Gauss-Wiener process with
variance $t$. The parameters $u$ and $\sigma$ are the drift and
the volatility of the return, respectively. At this stage, we set
these parameters as constants.

Given the risk-free rate $r$, and defining the market price of risk:
\[
\lambda=\frac{\mu-r}{\sigma}
\]

\noindent we can describe the diffusion process under the unique
risk-neutral measure\footnote{Also called equivalent martingale measure (EMM)}
($\mathbb{Q}$-measure) instead of the physical measure ($\mathbb{P}$-measure)
using the Girsanov's theorem (see \cite{sundaram1997equivalent} for
a detail explanation). In short, we define a new Brownian motion under
the Martingale measure of the form:

\[
\mathrm{d}W_{t}=\lambda\mathrm{d}t+\mathrm{d}\tilde{W}_{t}
\]

\noindent and replacing into Eq. \ref{eq:GBM}, the price dynamics
is described under the risk neutral measure, and it is given by\footnote{The Girsanov's theorem ensure a equivalent measure in which $W_{t}$
is a Wiener process and $S_{t}$ is a martingale (risk-neutral)}:

\begin{equation}
\frac{\mathrm{d}S_{t}}{S_{t}}=r\mathrm{d}t+\sigma\mathrm{d}\tilde{W}_{t}\label{eq:riskfree}
\end{equation}

By Itô's calculus, is possible to rewrite the Eq. \ref{eq:GBM} into:

\begin{equation}
\mathrm{d}\left(\ln S_{t}\right)=\left(r-\frac{\sigma^{2}}{2}\right)\mathrm{d}t+\sigma\mathrm{d}W_{t}\label{eq:Dlns}
\end{equation}

\noindent and labeling $x_{t}=\ln S_{t}$:

\begin{equation}
\mathrm{d}x_{t}=\left(r-\frac{\sigma^{2}}{2}\right)\mathrm{d}t+\sigma\mathrm{d}W_{t}\label{eq:Dlns-1}
\end{equation}
\\

The probability density $P(x_{t},t,x',t')$ for the random variable
$x_{t}$ evolves according to the Fokker-Planck (or forward Kolmogorov)
equation \cite{Risken1984}:

\begin{eqnarray}
\frac{\partial P}{\partial t} & = & -\frac{\partial}{\partial x}\left[\left(r-\frac{\sigma^{2}}{2}\right)P\right]+\frac{1}{2}\frac{\partial^{2}}{\partial x^{2}}\left(\sigma^{2}P\right)\nonumber \\
 & = & \frac{1}{2}\sigma^{2}\frac{\partial^{2}P}{\partial x^{2}}-\left(r-\frac{\sigma^{2}}{2}\right)\frac{\partial P}{\partial x}\label{eq:F-P}
\end{eqnarray}

\noindent with initial condition: 

\[
P\left(x_{t},t=0\right)=\delta(x)
\]
\\

Using the following simple transformation:

\[
c=e^{-rt}P
\]

\noindent and rewriting $x$ in terms of $S$ $(x=\ln S)$, the Eq.
\ref{eq:F-P} yields to the Black-Scholes equation in it standard
form \cite{black1973pricing}:

\begin{equation}
\frac{\partial c}{\partial t}=\frac{1}{2}\sigma^{2}S^{2}\frac{\partial^{2}c}{\partial S^{2}}+rs\frac{\partial c}{\partial x}-rc\label{eq:BS}
\end{equation}

Using the wick rotation ($\tilde{t}=it$), the evolution of the probability
density $P$ (Eq. \ref{eq:F-P}) can be mapped to the Sch\"odringer
equation:

\begin{equation}
i\hbar\frac{\partial\Psi}{\partial\tilde{t}}=\mathrm{\hat{H}}_{BS}\Psi\label{eq: Schrodinguer BS}
\end{equation}

\noindent where the wave function $\Psi$ represents the probability
$P$, and the quantum Hamiltonian $\hat{\mathrm{H}}_{BS}$, namely
for this instance the Black-Scholes Hamiltonian, is given by \cite{baaquie2004hamiltonian}:

\[
\hat{\mathrm{H}}_{BS}=\frac{1}{2}\sigma^{2}\frac{\partial^{2}}{\partial x^{2}}-\left(r-\frac{\sigma^{2}}{2}\right)\frac{\partial}{\partial x}
\]
\\

In order to ensure the compatibility between the Eqs. \ref{eq:F-P}
and \ref{eq: Schrodinguer BS}, we need to set $\hbar=1$. 

Given the momentum operator ${\displaystyle \mathrm{\hat{p}}=-i\hbar\frac{\partial}{\partial x}=-i\frac{\partial}{\partial x}}$,
the Hamiltonian can be expressed as :

\[
\hat{\mathrm{H}}_{BS}=-\frac{1}{2}\sigma^{2}\mathrm{\hat{p}}^{2}-i\left(r-\frac{\sigma^{2}}{2}\right)\mathrm{\hat{p}}
\]
\\

Considering $\Psi_{0}(x)$ as the initial value of $\Psi$ (i.e.,
$\Psi(x,t=0)=\Psi_{0}(x)$), the general solution of \ref{eq: Schrodinguer BS}
is given by (see \cite{baaquie2007quantum}):

\begin{eqnarray*}
\Psi(x,t) & = & e^{-i\hat{H}_{BS}\tilde{t}}\Psi_{0}(x)\\
 & = & e^{\hat{H}_{BS}t}\Psi_{0}(x)
\end{eqnarray*}
\\

However, the known conditions in the option pricing context (i.e.,
contract function) is set at time $T$. Thus, defining the backward
time $\tau=T-t$ and considering a final term value of the wave function
$\Psi(x,T)=\Psi_{T}(x_{T})$, the solution becomes:

\begin{equation}
\Psi(x,t)=e^{-\hat{H}_{BS}\tau}\Psi_{T}(x)\label{eq:sol tau}
\end{equation}
\\

Equivalently, using the convolution properties, the value at time
$t$ of the wave function is represented by:

\begin{eqnarray*}
\Psi(x,t) & = & \int_{-\infty}^{\infty}e^{-\hat{H}_{BS}\tau}\delta\left(x-x_{T}\right)\Psi_{T}\left(x_{T}\right)\mathrm{d}x_{T}\\
 & = & \int_{-\infty}^{\infty}K_{BS}\left(x,\tau|x_{T},0\right)\Psi_{T}\left(x_{T}\right)\mathrm{d}x_{T}
\end{eqnarray*}
\\

\noindent where $K_{BS}\left(x,\tau|x',0\right)$ is the propagator,
which admits the following path integral representation in euclidean
time \cite{feynman2010quantum}:

\[
K_{BS}\left(x,\tau|x_{T},0\right)=\int\mathcal{D}x(\tau)e^{-S_{BS}\left[x(\tau)\right]}
\]
\\

\noindent being $S_{BS}\left[x(\tau)\right]$ the euclidean classical
action along all the paths $x\left(t\right)$ which link the points
$x\left(T\right)=x_{T}$ and $x(t)=x$; defined by:

\[
S\left[x(t)\right]=\int_{t}^{T}\mathcal{L}_{BS}d\tau'
\]
\\

\noindent with $\mathcal{L}_{BS}$ the Lagrangian.

In order to obtain an expression for the propagator (Eqs. \pageref{eq:semi-1}-\pageref{eq:K semi-1})
we request the classical action evaluated over the classical path.
This can be obtained using the classical Hamiltonian mechanics.

The classical Hamiltonian $\mathcal{H}_{BS}$ associated to the operator
$\hat{\mathrm{H}}_{BS}$ is:

\[
\mathcal{H}_{BS}=-\frac{1}{2}\sigma^{2}p^{2}-i\left(r-\frac{\sigma^{2}}{2}\right)p
\]
\\

\noindent with its related classical Hamilton's equations in euclidean
time:

\begin{eqnarray*}
-i\dot{x} & = & \frac{\partial\mathcal{H}_{BS}}{\partial p}\\
-i\dot{p} & =- & \frac{\partial\mathcal{H}_{BS}}{\partial x}
\end{eqnarray*}
\\

\noindent or explicitly:

\begin{eqnarray}
p & = & \frac{i}{\sigma^{2}}\left[\dot{x}-\left(r-\frac{\sigma^{2}}{2}\right)\right]\label{eq:x_dot}\\
\dot{p} & = & 0\label{eq:p_dot}
\end{eqnarray}
\\

Then, the Lagrangian is given via the Legendre transformation:

\begin{eqnarray*}
\mathcal{L}_{BS} & = & -ip\dot{x}-\mathcal{H}_{BS}\\
 & = & -ip\dot{x}+\frac{1}{2}\sigma^{2}p^{2}+i\left(r-\frac{\sigma^{2}}{2}\right)p\\
 & = & \frac{p}{2}\left[\sigma^{2}p-2i\left(\dot{x}-r+\frac{\sigma^{2}}{2}\right)\right]
\end{eqnarray*}
\\

Using the values that solves the Hamilton's equation (Eqs. \ref{eq:x_dot}-\ref{eq:p_dot}),
the Lagrangian is:

\begin{equation}
\mathcal{L}_{BS}=\frac{1}{2\sigma^{2}}\left[\dot{x}-\left(r-\frac{\sigma^{2}}{2}\right)\right]^{2}\label{eq:BS-Lag}
\end{equation}
\\

Later, the Euler-Lagrange equation:

\begin{equation}
\frac{\mathrm{d}}{\mathrm{d}t}\left(\frac{\partial\mathcal{L}_{BS}}{\partial\dot{x}}\right)-\frac{\partial\mathcal{L}_{BS}}{\partial x}=0\label{eq:E-L eq}
\end{equation}
\\

\noindent yields to the free particle Newton equation:

\begin{equation}
\ddot{x}=0\label{eq:Newton-BS}
\end{equation}
\\

\noindent which leads to:

\begin{eqnarray}
\dot{x} & = & C\label{eq:xdot_int}\\
x & = & Ct+D\label{eq:integ_BS}
\end{eqnarray}
\\

The values for $C$ and $D$ are obtained using the border conditions
(fixed values) for $x$:

\begin{eqnarray*}
x(0) & = & x_{T}\\
x(\tau) & = & x_{0}
\end{eqnarray*}
\\

Thus, the classical path, with $0\leq\tau'\leq T$, is described by:

\begin{eqnarray}
\dot{x}(t) & = & \frac{x_{T}-x_{0}}{\tau}\label{eq:xdot_int-1}\\
x(t) & = & \frac{x_{T}-x_{0}}{\tau}\tau'+x_{0}\label{eq:integ_BS-1}
\end{eqnarray}
\\

Then, using \ref{eq:xdot_int-1} and \ref{eq:integ_BS-1}, the corresponding
classical action over the classical path:

\begin{eqnarray}
A\left[x_{class}\left(t\right)\right] & = & \int_{0}^{\tau}\frac{1}{2\sigma^{2}}\left[\dot{x}\left(\tau'\right)-\left(r-\frac{\sigma^{2}}{2}\right)\right]^{2}\mathrm{d}\left(\tau'\right)\nonumber \\
 & = & \frac{1}{2\sigma^{2}\tau}\left[x_{T}-x_{0}-\tau\left(r-\frac{\sigma^{2}}{2}\right)\right]^{2}\label{eq:A_class}
\end{eqnarray}
\\

Now, we are in conditions to compute the propagator. According to
Eqs. \ref{eq:normalization} and \ref{eq:vanVleck-1}, for this case:

\[
\mathcal{N}=\frac{1}{\sqrt{2\pi\sigma^{2}\left(T-t\right)}}
\]
\\

\noindent and the semiclassical approximation for the propagator
becomes:

\begin{eqnarray*}
K_{BS}^{SC}\left(x,\tau|x_{T},0\right) & = & \frac{e^{-S_{BS}\left[x_{class}(\tau)\right]}}{\sqrt{2\pi\sigma^{2}\left(T-t\right)}}\\
 & = & \frac{1}{\sqrt{2\pi\sigma^{2}\tau}}e^{-\frac{1}{2\sigma^{2}\tau}\left[x_{T}-x_{0}-\tau\left(r-\frac{\sigma^{2}}{2}\right)\right]^{2}}
\end{eqnarray*}
\\

Then, the wave function solution is reduced to:

\begin{eqnarray}
\Psi(x,t) & = & \frac{1}{\sqrt{2\pi\sigma^{2}\left(T-t\right)}}\int_{-\infty}^{\infty}e^{-S_{BS}\left[x_{class}(\tau)\right]}\Psi_{T}\left(x_{T}\right)\mathrm{d}x_{T}\label{eq:sol}\\
 & = & \frac{1}{\sqrt{2\pi\sigma^{2}\tau}}\int_{-\infty}^{\infty}e^{-\frac{1}{2\sigma^{2}\left(T-t\right)}\left[x_{T}-x_{0}-\left(T-t\right)\left(r-\frac{\sigma^{2}}{2}\right)\right]^{2}}\Psi_{T}\left(x_{T}\right)\mathrm{d}x_{T}
\end{eqnarray}

\noindent which is equal to the convolution between the propagator
and the contract function:

\[
\Psi\left(x,t\right)=K_{SC}^{BS}\ast\Psi_{T}\left(x_{T}\right)
\]
\\

The solution of Eq. \ref{eq:sol} depends on the border condition
$\Psi_{T}$ (contract function). We analyze the case of an European
call option, i.e.,:

\begin{eqnarray*}
\Psi_{T}\left(x_{T}\right) & = & e^{-r\tau}\max\left\{ S_{T}-E,0\right\} \\
 & = & e^{-r\tau}\max\left\{ e^{x_{T}}-E,0\right\} 
\end{eqnarray*}

\noindent being $E$ the strike price. \\

Then, the wave function for this case is:

\begin{eqnarray*}
\Psi(x,t) & = & \frac{e^{-r\tau}}{\sqrt{2\pi\sigma^{2}\tau}}\int_{\ln K}^{\infty}e^{-\frac{1}{2\sigma^{2}\tau}\left[x_{T}-x_{0}-\tau\left(r-\frac{\sigma^{2}}{2}\right)\right]^{2}}\left(e^{x_{T}}-E\right)\mathrm{d}x_{T}\\
 & = & \frac{e^{-r\tau}}{\sqrt{2\pi\sigma^{2}\tau}}\int_{\ln E}^{\infty}e^{x_{T}}e^{-\frac{1}{2\sigma^{2}\tau}\left[x_{T}-x_{0}-\tau\left(r-\frac{\sigma^{2}}{2}\right)\right]^{2}}\mathrm{d}x_{T}\\
 &  & -\frac{Ee^{-r\tau}}{\sqrt{2\pi\sigma^{2}\tau}}\int_{\ln E}^{\infty}e^{-\frac{1}{2\sigma^{2}\tau}\left[x_{T}-x_{0}-\tau\left(r-\frac{\sigma^{2}}{2}\right)\right]^{2}}\mathrm{d}x_{T}\\
 & = & I_{1}-I_{2}
\end{eqnarray*}
\\

Developing $I_{1},$ we have:

\begin{eqnarray*}
I_{1} & = & \frac{e^{-r\tau}}{\sqrt{2\pi\sigma^{2}\tau}}\int_{\ln E}^{\infty}e^{-\frac{1}{2\sigma^{2}\tau}\left[x_{T}^{2}-2x_{T}\left(x_{0}+\tau\left(r-\frac{\sigma^{2}}{2}\right)\right)+\left(x_{0}+\tau\left(r-\frac{\sigma^{2}}{2}\right)\right)^{2}\right]+x_{T}}\mathrm{d}x_{T}\\
 & = & \frac{e^{x_{0}}}{\sqrt{2\pi\sigma^{2}\tau}}\int_{\ln E}^{\infty}e^{-\frac{1}{2\sigma^{2}\tau}\left[x_{T}-\left(x_{0}+\tau\left(r+\frac{\sigma^{2}}{2}\right)\right)\right]^{2}}\mathrm{d}x_{T}
\end{eqnarray*}
\\

Carrying out the change of variable $u=\left[-x_{T}+x_{0}+\tau\left(r+\frac{\sigma^{2}}{2}\right)\right]/\sqrt{\sigma^{2}\tau}$,
and replacing $x_{0}=\ln S_{0}$, we have:

\begin{eqnarray*}
I_{1} & =- & S_{0}\left[\int_{x_{0}-\ln E+\tau\left(r+\frac{\sigma^{2}}{2}\right)}^{-\infty}e^{-\frac{1}{2}v^{2}}\mathrm{d}u\right]\\
 & = & S_{0}N\left(d_{1}\right)
\end{eqnarray*}
\\

\noindent where $N\left(\cdot\right)$ is the standard normal cumulative
function and $d_{1}=\frac{\ln\left(\frac{S_{0}}{E}\right)+\tau\left(r+\frac{\sigma^{2}}{2}\right)}{\sqrt{\sigma^{2}\text{\ensuremath{\tau}}}}$
.

For to solve $I_{2}$ we use the change of variable $v=-\left[x_{T}-x_{0}-\tau\left(r-\frac{\sigma^{2}}{2}\right)\right]/\sqrt{\sigma^{2}\tau}$,
so:

\begin{eqnarray*}
I_{2} & = & -\frac{Ee^{-r\tau}}{\sqrt{2\pi}}\int_{x_{0}-\ln E+\tau\left(r-\frac{\sigma^{2}}{2}\right)}^{-\infty}e^{-\frac{1}{2}v^{2}}\mathrm{d}u\\
 & = & Ke^{-r\tau}N\left(d_{2}\right)
\end{eqnarray*}
\\

\noindent being $d_{2}=\frac{\ln\left(\frac{S_{0}}{E}\right)+\tau\left(r-\frac{\sigma^{2}}{2}\right)}{\sqrt{\sigma^{2}\text{\ensuremath{\tau}}}}=d_{1}-\sqrt{\sigma^{2}\text{\ensuremath{\tau}}}$
.\\

Finally, the price of a call option at time $t$, using the path integral
formul\ae is given by:

\[
\Psi\left(S_{0},t\right)=S_{0}N\left(d_{1}\right)-Ke^{-r\tau}N\left(d_{2}\right)
\]
\\

\noindent which is exactly the same value obtained by Black-Scholes
\cite{black1973pricing} for an European call option\footnote{As noted previously (On page \pageref{quadratic}), the semiclassical
approximation is exact if the Lagrangian is quadratic, as in the case
of the B-S Lagrangian (Eq. \ref{eq:BS-Lag}) }.

\section{A semiclassical approximation of the path integral approach to the
CEV model}

In the CEV model, under the risk-neutral measure, the asset is governed
by the following stochastic differential equation \cite{Cox1975,cox1996constant}:\\

\begin{equation}
\mathrm{d}S\left(S,t\right)=rS\mathrm{d}t+\sigma S^{\frac{\alpha}{2}}\mathrm{d}W\label{eq:CEV}
\end{equation}

\noindent being $r$ the constant risk-free of interest, $\sigma$and
$\alpha$ taking constant values and $W$ a standard Wiener process,
whit d$W\sim N\left(0,\mathrm{d}t\right)$. In its paper, Cox imposed
the domain for $\alpha$ in the range $[0,2[$. In this interval the
negative relationship between the asset level and volatility is observed
(leverage effect). For values greater than two, the process described
in the Eq. \ref{eq:CEV} is not a martingale \cite{veestraeten2017multiplicity,delbaen2002note}
(i.e, there are not a unique risk-neutral measure). For $\alpha<0$,
the volatility unrealistically goes to zero as $S$ increases \cite{lorig2013exact}.
Then, the same Cox's condition for $\alpha$ is assumed in this paper.

The process described by the Eq. \ref{eq:CEV} can be interpreted
as a generalization of the standard geometric Brownian motion used
in the Black-Scholes model \cite{black1973pricing}, but considering
a non-constant local volatility function equals to $\sigma S^{\frac{\alpha-2}{2}}$.
In fact, for the limit case $\alpha=2$, the Eq. \ref{eq:CEV} is
degenerated to the BS case. Also, the CEV model has correspondence
with other approaches: For $\alpha=1$, it becomes a square root process,
addressed by Cox and Ross \cite{cox1976valuation}; and for $\alpha=0$
, $S$ follows an Ornstein-Uhlenbeck type process \cite{uhlenbeck1930theory}.

The CEV model described in Eq. \ref{eq:CEV} owes its name to the
fact that the variance of the return is given by:

\begin{eqnarray*}
v & = & \mathrm{var}\left(\frac{\mathrm{d}S}{S}\right)\\
 & = & \mathrm{var}\left(r\mathrm{d}t+\sigma S^{\frac{\alpha-2}{2}}\mathrm{d}W\right)\\
 & = & \sigma^{2}S^{\alpha-2}\mathrm{d}t
\end{eqnarray*}

\noindent and then, the elasticity of the variance with respect to
the spot:

\[
\frac{\mathrm{d}v/v}{\mathrm{d}S/S}=\alpha-2
\]

\noindent is constant.

The strategy to get an option pricing formula will be the same that
it was developed in section \ref{sec:The-Black-Scholes}. That is:
i) we arrive to the Fokker-Planck equation; ii) we rewrite it as a
Schr\"{o}dinger equation; iii) later, we find the classical path
through the Hamilton or Euler-Lagrange equations, working with the
propagator as path integral, iv) we evaluate the classical path using
semiclassical arguments; and v) finally, we compute the convolution
between the propagator and the contract function in the integral form.\\

Firstly, we use the following transformation:

\[
y(S,t)=S^{2-\alpha}
\]

\noindent and by the Itô's Lemma, Eq. \ref{eq:CEV} can be rewrite
as:

\[
\mathrm{d}y=\left(2-\alpha\right)\left[ry+\frac{1}{2}\left(1-\alpha\right)\sigma^{2}\right]\mathrm{d}t+\left(2-\alpha\right)\sigma\sqrt{y}\mathrm{d}W
\]
\\

The Fokker-Planck equation rules the transition probability $P(Y,t)$
of the variable $Y$. Thus:

\begin{eqnarray}
\frac{\partial P}{\partial t} & = & \frac{1}{2}\frac{\partial^{2}}{\partial y^{2}}\left[\left(2-\alpha\right)^{2}\sigma^{2}yP\right]-\frac{\partial}{\partial y}\left[\left(2-\alpha\right)\left(ry+\frac{1}{2}\left(1-\alpha\right)\sigma^{2}\right)P\right]\nonumber \\
 & = & \frac{1}{2}\beta^{2}\sigma^{2}y\frac{\partial^{2}P}{\partial y^{2}}+\beta r\left[\gamma-y\right]\frac{\partial P}{\partial y}-\beta rP\label{eq:fokker-planck-CEV}
\end{eqnarray}
\\

\noindent being $\beta$ and $\gamma$ constant values (parameters),
defined as: 
\[
\beta=2-\alpha
\]

\[
\gamma=\frac{3-\alpha}{2r}\sigma^{2}
\]
\\

The relationship \ref{eq:fokker-planck-CEV} can be interpreted as
the Schr\"{o}dinger equation in Euclidean (Wick-rotated) time, with
$\hbar=1$:

\[
\frac{\partial\Psi}{\partial t}=\hat{\text{H}}\Psi
\]
\\

\noindent where the wave function $\Psi$ is equivalent to the probability
$P$ and the Hamiltonian operator $\hat{\text{H}}$ is given by:

\[
\hat{\text{H}}=\frac{1}{2}\beta^{2}\sigma^{2}y\frac{\partial^{2}}{\partial y^{2}}+\beta r\left[\gamma-y\right]\frac{\partial}{\partial y}-\beta r
\]
\\

Using the quantum momentum operator, $\hat{p}=-i\frac{\partial}{\partial Y}$,
the Hamiltonian goes to:

\[
\hat{\text{H}}=-\frac{1}{2}\beta^{2}\sigma^{2}y\hat{p}^{2}+i\beta r\left[\gamma-y\right]\hat{p}-\beta r
\]
\\

Later, we consider a final term condition (contract function) of the
form:

\[
\Psi(y,t=T)=\Psi(y_{T})
\]
\\

The wave function $\Psi$, can be written in terms of it propagator
$K$:

\[
\Psi\left(y,t\right)=\int_{-\infty}^{\infty}K\left(y,\tau|y_{T},0\right)\text{\ensuremath{\Psi}}(y_{T})\mathrm{d}y_{T}
\]
\\

\noindent where $\text{\ensuremath{\tau}}=T-t$, is the backward
time and: 

\[
K\left(y,\tau|y_{T},0\right)=<y|e^{-\mathrm{\tau\hat{H}}}|y_{T}>=e^{-\hat{H}\tau}\delta\left(y-y_{T}\right)
\]
\\
On the other hand, the propagator can be estimated using the path
integral:

\[
K\left(y_{T},T|y_{0},0\right)=\int\mathcal{D}\left[y(t)\right]e^{-S\left[y(t)\right]}
\]
\\

\noindent being $\mathcal{D}\left[y(\tau)\right]$ the infinitesimal
contribution of all the paths $y(\tau)$ that satisfies the boundary
conditions $y(t=T)=y_{T}$ and $y(t=0)=y_{0}$; and $S$ the euclidean
classical action functional over $y\left(t\right)$.

Using semiclassical arguments the propagator becomes:

\[
K\left(y_{T},0|y_{0},\tau\right)=e^{-A\left[y_{class}\left(t\right)\right]}\sqrt{\frac{1}{2\pi}\mathcal{M}}
\]
\\

The classical path is obtained as the solution of the Hamilton equations.
The classical Hamiltonian $\mathcal{H}$ related to $\hat{\mathrm{H}}$
is:

\[
\mathcal{H}=-\frac{1}{2}\beta^{2}\sigma^{2}yp^{2}+i\beta r\left[\gamma-y\right]p-\beta r
\]
\\

\noindent where $p$ represents the classical momentum. Considering
the Hamilton equation in Euclidean time, the momentum can be written
in terms of $y$ and $\dot{y}$:

\begin{equation}
p=i\frac{\dot{y}+\beta r\left[\gamma-y\right]}{\beta^{2}\sigma^{2}y}\label{eq:p_classic}
\end{equation}
\\

So, using Eq. \ref{eq:p_classic}, the Lagrangian takes the form:

\begin{eqnarray}
\mathcal{L} & = & -i\dot{y}p-\mathcal{H}\label{eq:Lagrangiano}\\
 & = & \frac{\left\{ \dot{y}+\beta r\left[\gamma-y\right]\right\} ^{2}}{2\mathbf{\beta}^{2}\sigma^{2}y}+Ar\nonumber 
\end{eqnarray}
\\

The unique classical trajectory is which obeys the Euler-Lagrange
equation:

\begin{equation}
\frac{\text{d}}{\text{d}t}\left(\frac{\partial\mathcal{L}}{\partial\dot{y}}\right)-\frac{\partial\mathcal{L}}{\partial y}=0\label{eq:E-L-gen}
\end{equation}
\\

Computing the derivatives:

\begin{eqnarray*}
\frac{\partial\mathcal{L}}{\partial\dot{y}} & = & \frac{\dot{y}+\beta r\left(\gamma-y\right)}{\beta^{2}\sigma^{2}y}\\
\frac{\text{d}}{\text{d}t}\left(\frac{\partial\mathcal{L}}{\partial\dot{y}}\right) & = & \frac{\ddot{y}y-\dot{y}^{2}-\beta\gamma r\dot{y}}{\beta^{2}\sigma^{2}y^{2}}\\
\frac{\partial\mathcal{L}}{\partial y} & = & -\frac{\left[\dot{y}+\beta r\left(\gamma+y\right)\right]\left[\dot{y}+\mathbf{\beta}r\left(\gamma-y\right)\right]}{2\beta^{2}\sigma^{2}y^{2}}
\end{eqnarray*}
\\

\noindent and replacing into the Eq. \ref{eq:E-L-gen}, we have a
second order differential equation that rules the classical behavior
of $y(t)$:

\begin{equation}
2y\ddot{y}-\dot{y}^{2}+\beta^{2}r^{2}\left(\gamma^{2}-y^{2}\right)=0\label{eq:E-L}
\end{equation}
\\

Then, solving the Eq. \ref{eq:E-L}, the classical path is given by:

\begin{equation}
y_{class}\left(t\right)=\frac{\left(C_{1}+2C_{2}e^{-rt\beta}\right)^{2}-\gamma^{2}}{4C_{2}e^{-rt\beta}}\label{eq:classical path}
\end{equation}
\\

\noindent being $C_{1}$ and $C_{2}$ constants given by the fixed
values of the path at time $t=0$ and $t=T$:

\[
\begin{array}{c}
y(T)=y_{T}\\
y(t_{0})=y_{0}
\end{array}
\]
\\

\noindent which yields to:

\begin{equation}
C_{1}=\frac{\left(e^{r\tau\beta}+1\right)\sqrt{\gamma^{2}\left(e^{r\tau\beta}-1\right)^{2}+4y_{0}y_{T}e^{r\tau\beta}}-2e^{r\tau\beta}\left(y_{0}+y_{T}\right)}{\left(e^{r\tau\beta}-1\right)^{2}}\label{eq:c1}
\end{equation}

\begin{equation}
C_{2}=\frac{\left[y_{T}e^{r\tau\beta}+y_{0}-\sqrt{\gamma^{2}\left(e^{r\tau\beta}-1\right)^{2}+4y_{0}y_{T}e^{r\tau\beta}}\right]}{\left(e^{r\tau\beta}-1\right)^{2}}\label{eq:c2}
\end{equation}
\\

Later, using Eq. \ref{eq:classical path}, the Lagrangian over the
classical path is:

\begin{eqnarray}
\mathcal{L}_{class} & = & \mathcal{L}\left[y_{class}\right]\nonumber \\
 & = & \frac{r^{2}\left(C_{1}+2C_{2}e^{rt\beta}+\gamma\right)\left(\gamma-C_{1}\right)^{2}}{2\sigma^{2}C_{2}e^{rt\beta}\left(C_{1}+2C_{2}e^{rt\beta}-\gamma\right)}+\beta r\label{eq:L_class}
\end{eqnarray}
\\

Thus the classical action is obtained by time integration of the Eq.
\ref{eq:L_class} :

\begin{eqnarray}
A_{class} & = & \intop_{t=t_{0}}^{t=T}\mathcal{L}_{class}dt\nonumber \\
 & = & \left.\frac{r}{\sigma^{2}}\left\{ \beta\sigma^{2}t-2\gamma rt+\frac{2\gamma}{\beta}\ln\left[\gamma-\left(C_{1}+2C_{2}\text{e}^{rt\beta}\right)\right]+\frac{\left(\gamma^{2}-C_{1}^{2}\right)}{2A\beta\text{e}^{rt\beta}}\right\} \right|_{t=0}^{t=T}\nonumber \\
 & = & \beta r\tau-\frac{2\gamma r^{2}}{\sigma^{2}}\tau+\frac{2\gamma r}{\beta\sigma^{2}}\ln\left[\frac{\gamma-\left(C_{1}+2C_{2}\text{e}^{r\tau\beta}\right)}{\gamma-\left(C_{1}+2C_{2}\right)}\right]\nonumber \\
 &  & +\frac{\left(\gamma^{2}-C_{1}^{2}\right)}{2\beta C_{2}\text{e}^{r\tau\beta}}\left(1-\text{e}^{r\tau\beta}\right)\label{eq:Action}
\end{eqnarray}

So, using the Eq. \pageref{eq:Action}, the van Vleck determinant
(Eq. \ref{eq:vanVleck-1}) is computed as:

\begin{eqnarray*}
\mathcal{M} & = & \frac{2\gamma r\left[\gamma-\left(C_{1}+2C_{2}\right)\right]}{\beta\sigma^{2}\left[\gamma-\left(C_{1}+2C_{2}\text{e}^{r\tau\beta}\right)\right]}\Biggr\{\frac{-\frac{\partial^{2}}{\partial x_{0}\partial x_{T}}\left(C_{1}+2C_{2}\text{e}^{r\tau\beta}\right)}{\gamma-\left(C_{1}+2C_{2}\right)}\\
 &  & -\frac{\left[\frac{\partial}{\partial x_{0}}\left(C_{1}+2C_{2}\right)\right]\left[\frac{\partial}{\partial x_{T}}\left(C_{1}+2C_{2}\text{e}^{r\tau\beta}\right)\right]+\left[\frac{\partial}{\partial x_{T}}\left(C_{1}+2C_{2}\right)\right]\left[\frac{\partial}{\partial x_{0}}\left(C_{1}+2C_{2}\text{e}^{r\tau\beta}\right)\right]}{\left[\gamma-\left(C_{1}+2C_{2}\right)\right]^{2}}\\
 &  & +\frac{2\left[\gamma-\left(C_{1}+2C_{2}\text{e}^{r\tau\beta}\right)\right]\left[\frac{\partial}{\partial x_{T}}\left(C_{1}+2C_{2}\right)\right]\left[\frac{\partial}{\partial x_{0}}\left(C_{1}+2C_{2}\right)\right]}{\left[\gamma-\left(C_{1}+2C_{2}\right)\right]^{3}}\\
 &  & +\frac{\left[\gamma-\left(C_{1}+2C_{2}\text{e}^{r\tau\beta}\right)\right]\left[\frac{\partial^{2}}{\partial x_{0}\partial x_{T}}\left(C_{1}+2C_{2}\right)\right]}{\left[\gamma-\left(C_{1}+2C_{2}\right)\right]^{2}}\Biggr\}\\
 &  & -\Biggl\{\frac{2\gamma r\left[\gamma-\left(C_{1}+2C_{2}\right)\right]\left[-\frac{\partial}{\partial x_{0}}\left(C_{1}+2C_{2}\text{e}^{r\tau\beta}\right)\right]}{\beta\sigma^{2}\left[\gamma-\left(C_{1}+2C_{2}\text{e}^{r\tau\beta}\right)\right]^{2}}\\
 &  & \times\left[\frac{-\frac{\partial}{\partial x_{T}}\left(C_{1}+2C_{2}\text{e}^{r\tau\beta}\right)}{\gamma-\left(C_{1}+2C_{2}\right)}+\frac{\left[\gamma-\left(C_{1}+2C_{2}\text{e}^{r\tau\beta}\right)\right]\left[\frac{\partial}{\partial x_{T}}\left(C_{1}+2C_{2}\right)\right]}{\left[\gamma-\left(C_{1}+2C_{2}\right)\right]^{2}}\right]\Biggr\}\\
 &  & +\frac{2\gamma r\left[-\frac{\partial}{\partial x_{0}}\left(C_{1}+2C_{2}\right)\right]}{\beta\sigma^{2}\left[\gamma-\left(C_{1}+2C_{2}\text{e}^{r\tau\beta}\right)\right]}\Biggl\{\frac{\left[\gamma-\left(C_{1}+2C_{2}\text{e}^{r\tau\beta}\right)\right]\left[\frac{\partial}{\partial x_{T}}\left(C_{1}+2C_{2}\right)\right]}{\left[\gamma-\left(C_{1}+2C_{2}\right)\right]^{2}}\\
 &  & \frac{-\frac{\partial}{\partial x_{T}}\left(C_{1}+2C_{2}\text{e}^{r\tau\beta}\right)}{\gamma-\left(C_{1}+2C_{2}\right)}\Biggr\}+\frac{r\left(\text{e}^{\beta r\tau}-1\right)\left[\left(\frac{\partial C_{1}}{\partial x_{0}}\right)\left(\frac{\partial C_{1}}{\partial x_{T}}\right)+\frac{\partial^{2}C_{1}}{\partial x_{0}\partial x_{T}}\right]}{\beta\sigma^{2}C_{2}\text{e}^{r\tau\beta}}\\
 &  & -\frac{r\left(\text{e}^{\beta r\tau}-1\right)\left[\left(\frac{\partial C_{1}}{\partial x_{T}}\right)\left(\frac{\partial C_{2}}{\partial x_{0}}\right)+\left(\frac{\partial C_{1}}{\partial x_{0}}\right)\left(\frac{\partial C_{2}}{\partial x_{T}}\right)-\frac{1}{2}\left(\gamma^{2}-C_{1}^{2}\right)\left(\frac{\partial^{2}C_{2}}{\partial x_{0}\partial x_{T}}\right)\right]}{\beta\sigma^{2}C_{2}^{2}\text{e}^{r\tau\beta}}\\
 &  & -\frac{r\left(\text{e}^{\beta r\tau}-1\right)\left(\gamma^{2}-C_{1}^{2}\right)\left(\frac{\partial C_{2}}{\partial x_{T}}\right)\left(\frac{\partial C_{2}}{\partial x_{0}}\right)}{\beta\sigma^{2}C_{2}^{3}\text{e}^{r\tau\beta}}
\end{eqnarray*}

Then is possible to compute the semiclassical propagator, through
the Euclidean form of the Pauli's formula (Eq. \pageref{eq:K semi-1}):

\[
K=e^{-A\left[y_{class}\left(t\right)\right]}\sqrt{\frac{1}{2\pi}\mathcal{M}}
\]
\\

Finally, the value of the wave function at time $t$, is given by:

\[
\Psi\left(y,t\right)=\sqrt{\frac{1}{2\pi}}\int_{-\infty}^{\infty}\sqrt{\mathcal{M}}e^{-A_{class}}\text{\ensuremath{\Psi}}(y_{T})\mathrm{d}y_{T}
\]
\\

Coming back to the option pricing problem, if we consider an European
call option, with strike $E$ and maturity $T$, the value of the
option at time $t$ under the CEV model will be:

\begin{equation}
C\left(S,t\right)=\sqrt{\frac{1}{2\pi}}\int_{E^{1/(2-\alpha)}}^{\infty}\sqrt{\mathcal{M}}e^{-A\left[y_{class}\left(t\right)\right]}\left(y_{T}^{2-\alpha}-E\right)\mathrm{d}y_{T}\label{eq:Call-CEV}
\end{equation}
\\

\noindent which unfortunately is not possible to evaluate analytically,
but it can be easily computed numerically for any conventional integration
method.

\section{Numerical Simulations}

We compute numerically, using an standard method (global adaptive
quadrature \cite{shampine2008vectorized}), the integral defined in
Eq. \ref{eq:Call-CEV}. We also compute the pricing for the same European
call option using the Schroder approach \cite{schroder1989computing}
that consider the non-central chi-square distribution, and set it
as the benchmark.

We examine the results of both models, in terms of the pricing and
the running time of each computation; considering several volatilities
and elasticities of variances. Besides we test our results for short-time
maturities ($T=\{0.25,0.5\}$) and long time maturities ($T=\{2,4\}$).
In all the experiments we assume $r=0.05
$, $S_{0}=100$ and $E=110$ ,

Firstly, we consider a maturity equal to six months. In Table \ref{tab:T05}
both the pricing and computational time are reported. We can see that
the path integral method has similar pricing values but with a clear
advantage in the running time. The times observed in the Table \ref{tab:T05}
for the proposed method of path integral are always lower to 0.008
seconds; however for the non-central chi-square approach the times
are at least greater in one order of magnitude with a increase when
the elasticity parameter is higher. 

\begin{table}[H]
\begin{centering}
\begin{tabular}{|c|c|c|c|c|c|}
\hline 
\multirow{2}{*}{$\sigma$} &
\multirow{2}{*}{$\alpha$} &
\multicolumn{2}{c|}{Path Integral} &
\multicolumn{2}{c|}{Benchmark}\tabularnewline
\cline{3-6} 
 &  & Pricing ($\$$) &
Running Time (s) &
Pricing $(\$)$ &
Running Time (s)\tabularnewline
\hline 
\hline 
\multirow{3}{*}{$20\%$} &
1 &
4.4289e-08 &
0.0079 &
4.6567e-08 &
0.1595\tabularnewline
\cline{2-6} 
 & 1.45 &
0.0580 &
0.0064 &
0.0600 &
0.0886\tabularnewline
\cline{2-6} 
 & 1.9 &
1.8505 &
0.0060 &
1.8706 &
0.2101\tabularnewline
\hline 
\multirow{3}{*}{$50\%$} &
1 &
0.0259 &
0.0078 &
0.0583 &
0.0275\tabularnewline
\cline{2-6} 
 & 1.45 &
1.3437 &
0.0079 &
1.4181  &
0.0480\tabularnewline
\cline{2-6} 
 & 1.9 &
8.0777 &
0.0059 &
8.2636  &
0.0603\tabularnewline
\hline 
\multirow{3}{*}{$90\%$} &
1 &
0.3847 &
0.0059 &
0.4148 &
0.0307\tabularnewline
\cline{2-6} 
 & 1.45 &
3.9003 &
0.0077 &
4.2358 &
0.0236\tabularnewline
\cline{2-6} 
 & 1.9 &
16.4965 &
0.0074 &
17.1870 &
0.0413\tabularnewline
\hline 
\end{tabular}
\par\end{centering}
\caption{Comparison of pricing and computational time for a Call option for
some values of $\sigma$ and $\alpha$, using $T=0.5$, $S_{0}=100$
and $E=110$ \label{tab:T05}}
\end{table}

\begin{figure}[H]
\subfloat[Option Pricing]{\includegraphics[width=0.5\textwidth]{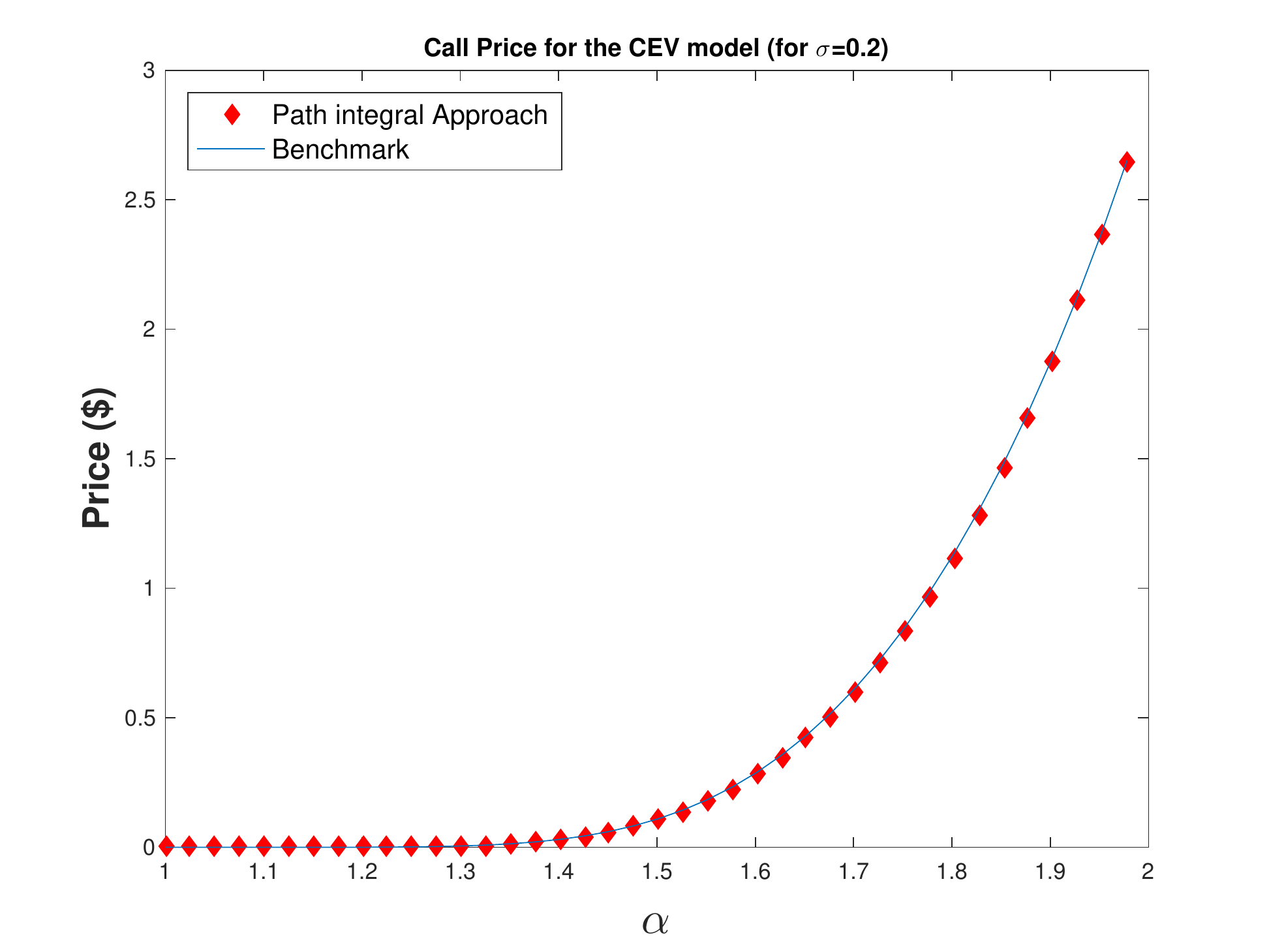}

}\subfloat[Running time]{\includegraphics[width=0.5\textwidth]{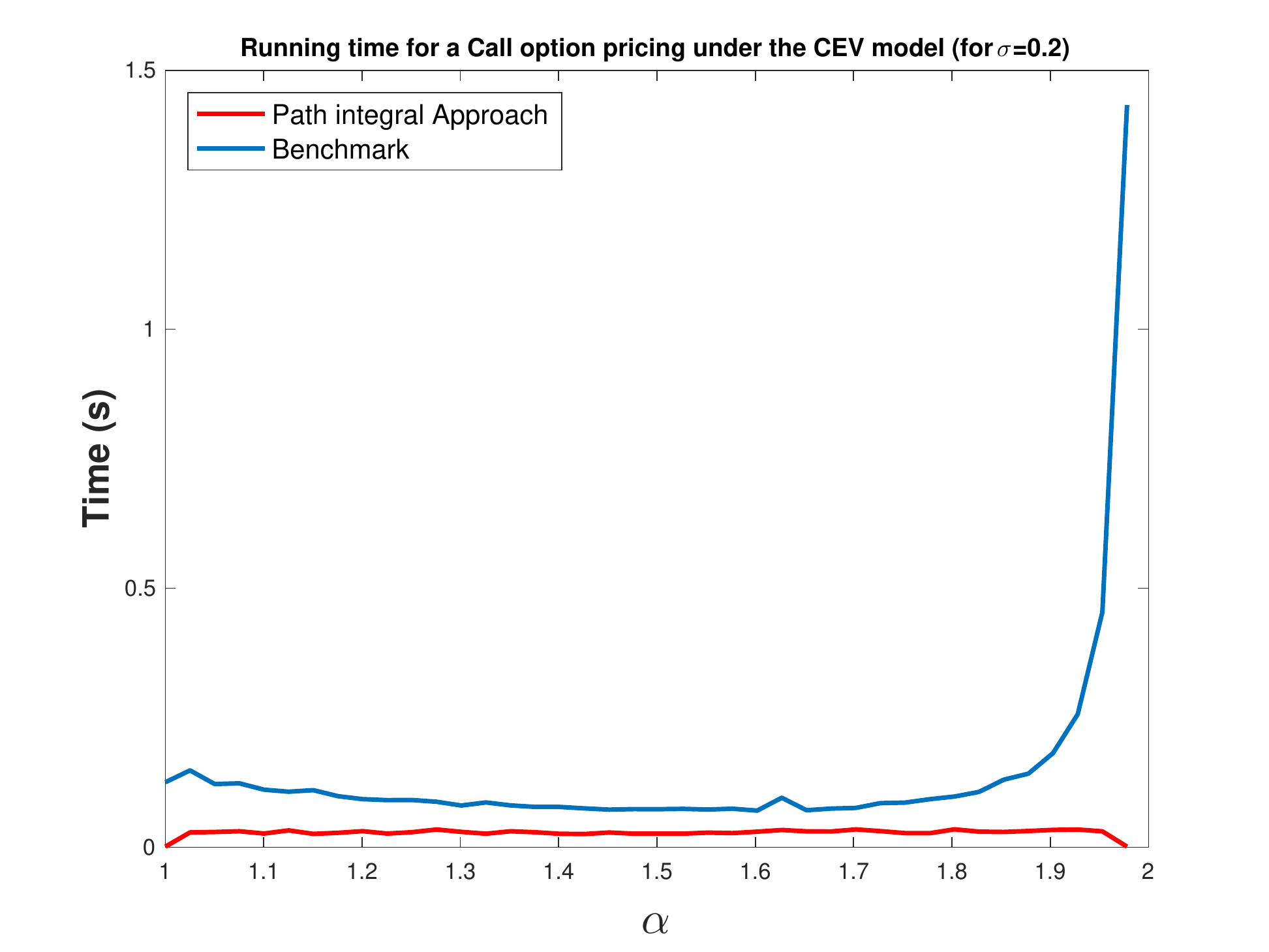}

}

\caption{Pricing and computational time for a Call option using $\sigma=$20\%,
$T=0.5$, $r=0.05$, $S_{0}=100$ and $E=110$ \label{fig:Comparisonsigma02T05}}
\end{figure}

\begin{figure}[H]
\subfloat[Option Pricing]{\includegraphics[width=0.5\textwidth]{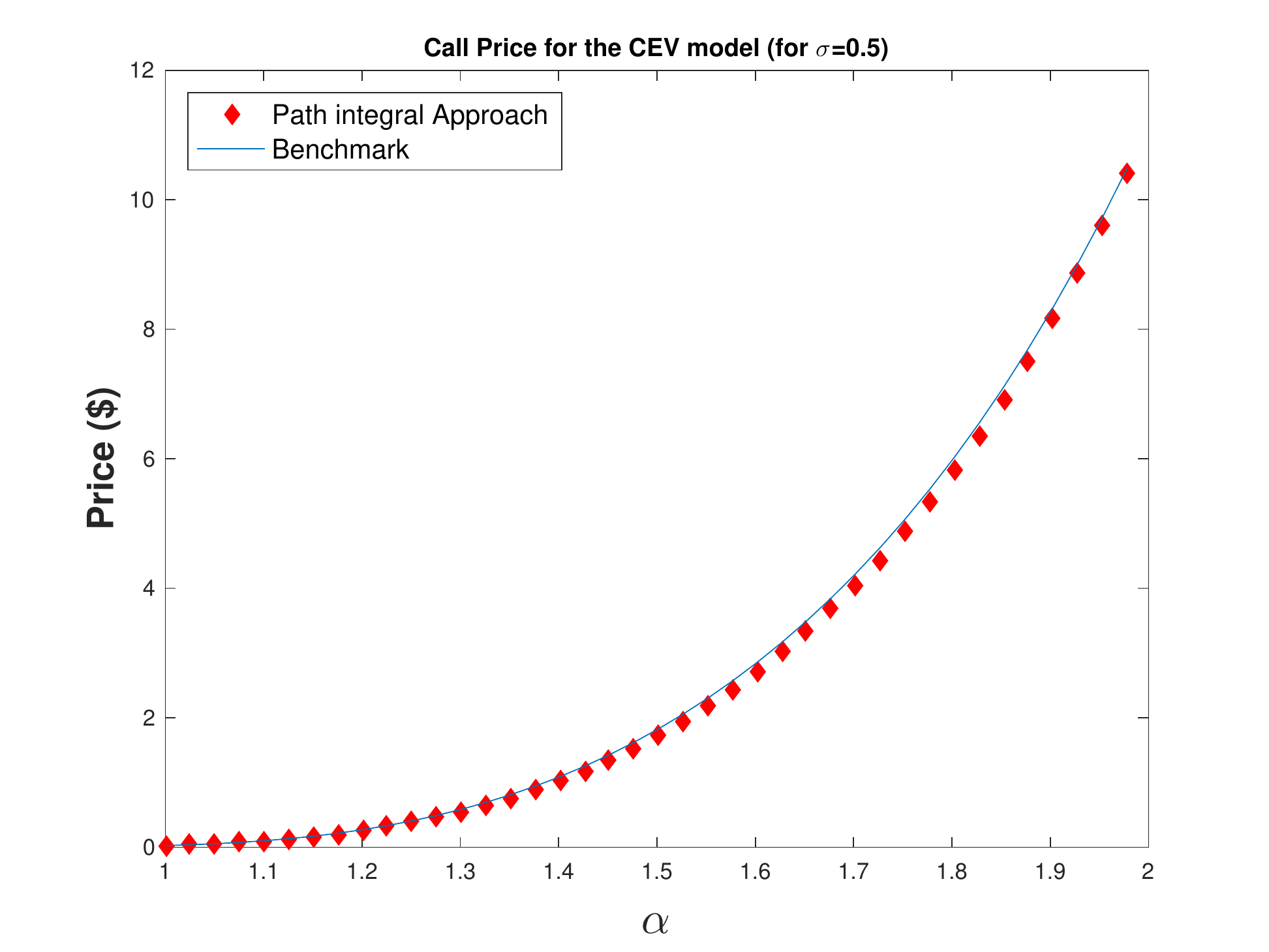}

}\subfloat[Running time]{\includegraphics[width=0.5\textwidth]{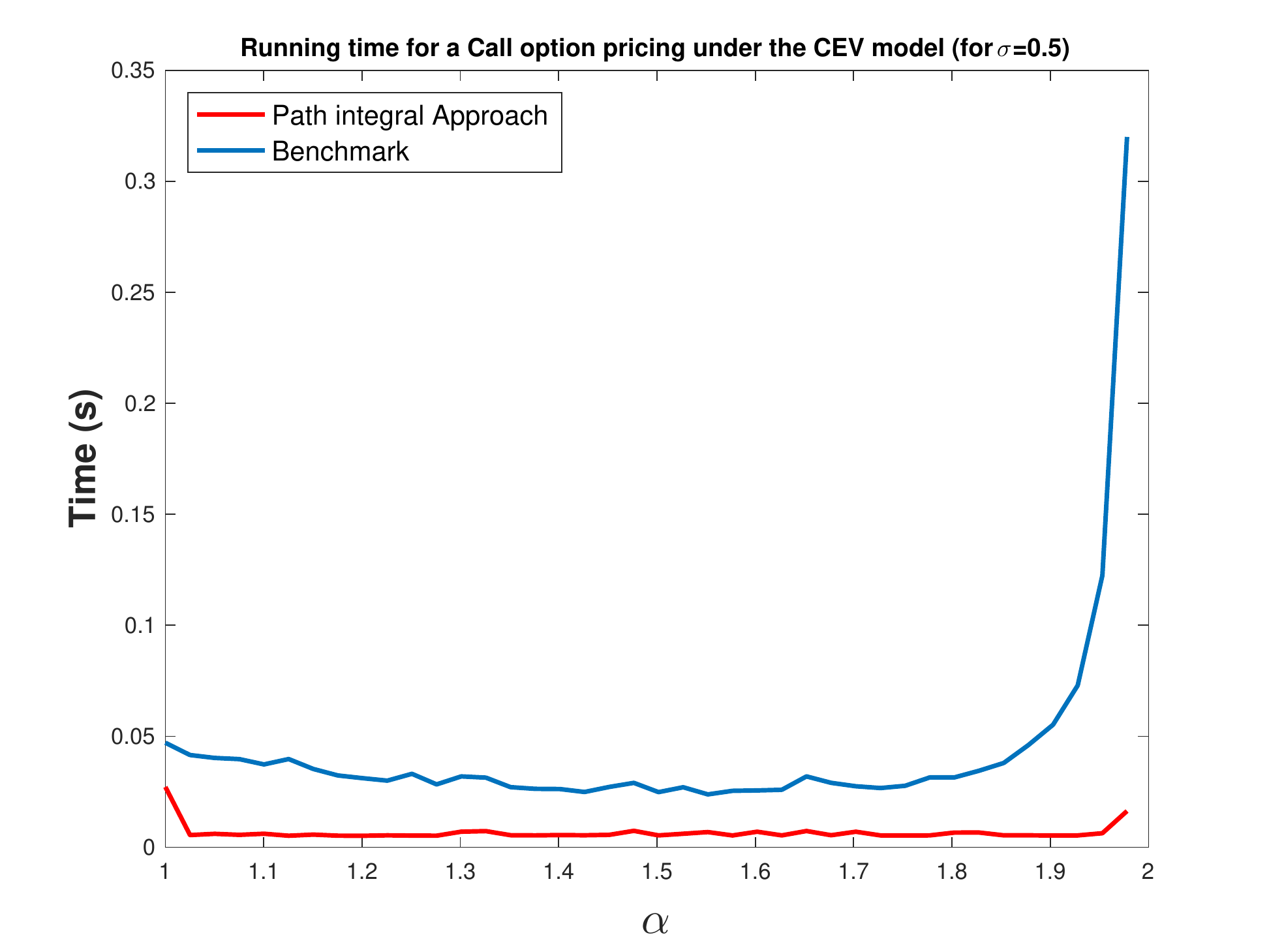}

}

\caption{Pricing and computational time for a Call option using $\sigma=$50\%,
$T=0.5$, $r=0.05$, $S_{0}=100$ and $E=110$\label{fig:Comparisonsigma5T05}}
\end{figure}

\begin{figure}[H]
\subfloat[Option Pricing]{\includegraphics[width=0.5\textwidth]{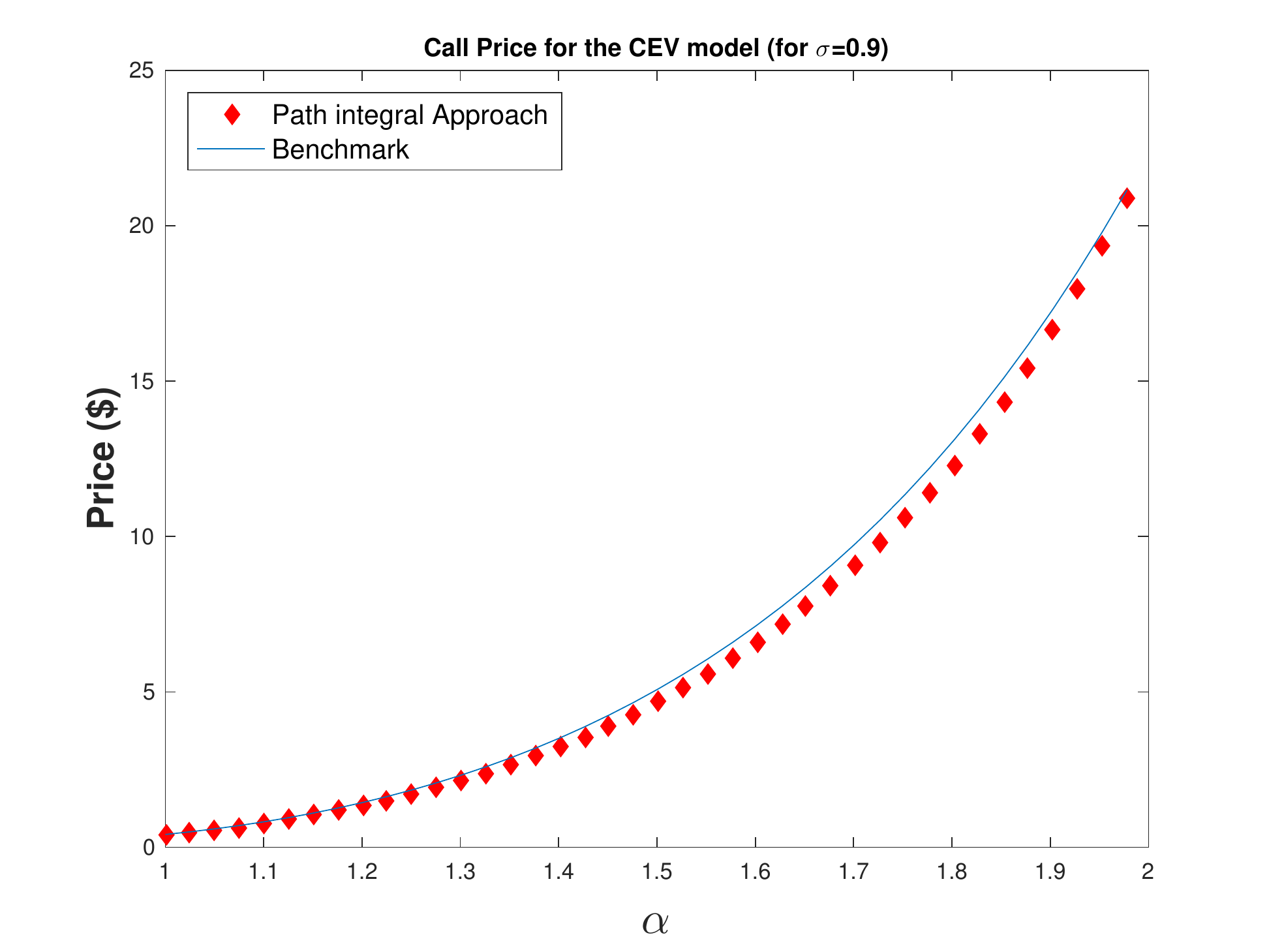}

}\subfloat[Running time]{\includegraphics[width=0.5\textwidth]{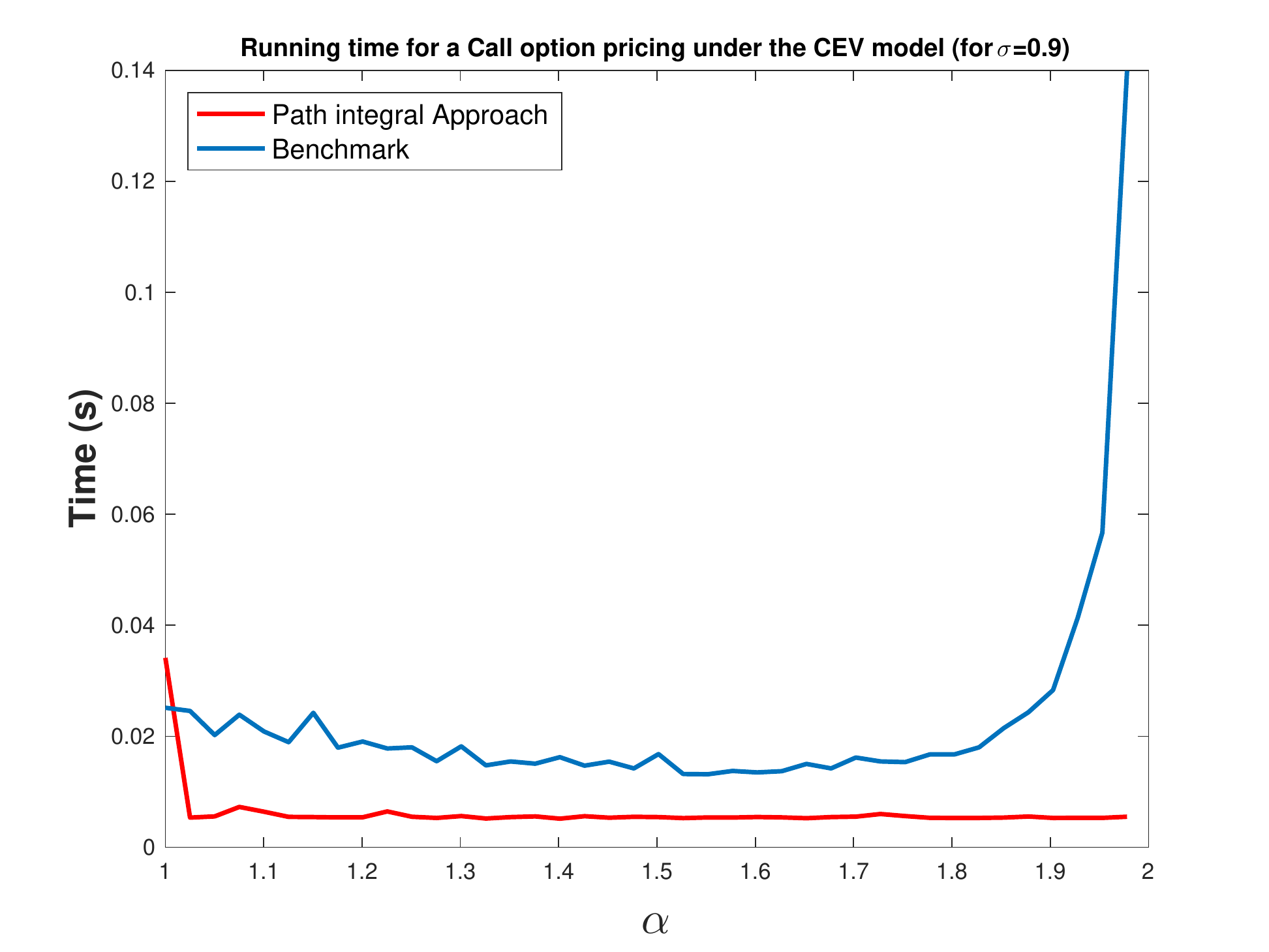}

}

\caption{Pricing and computational time for a Call option using $\sigma=$90\%,
$T=0.5$, $r=0.05$, $S_{0}=100$ and $E=110$ \label{fig:Comparisonsigma9T05}}
\end{figure}

\begin{figure}[H]
\subfloat[Absolute error]{\includegraphics[width=0.5\textwidth]{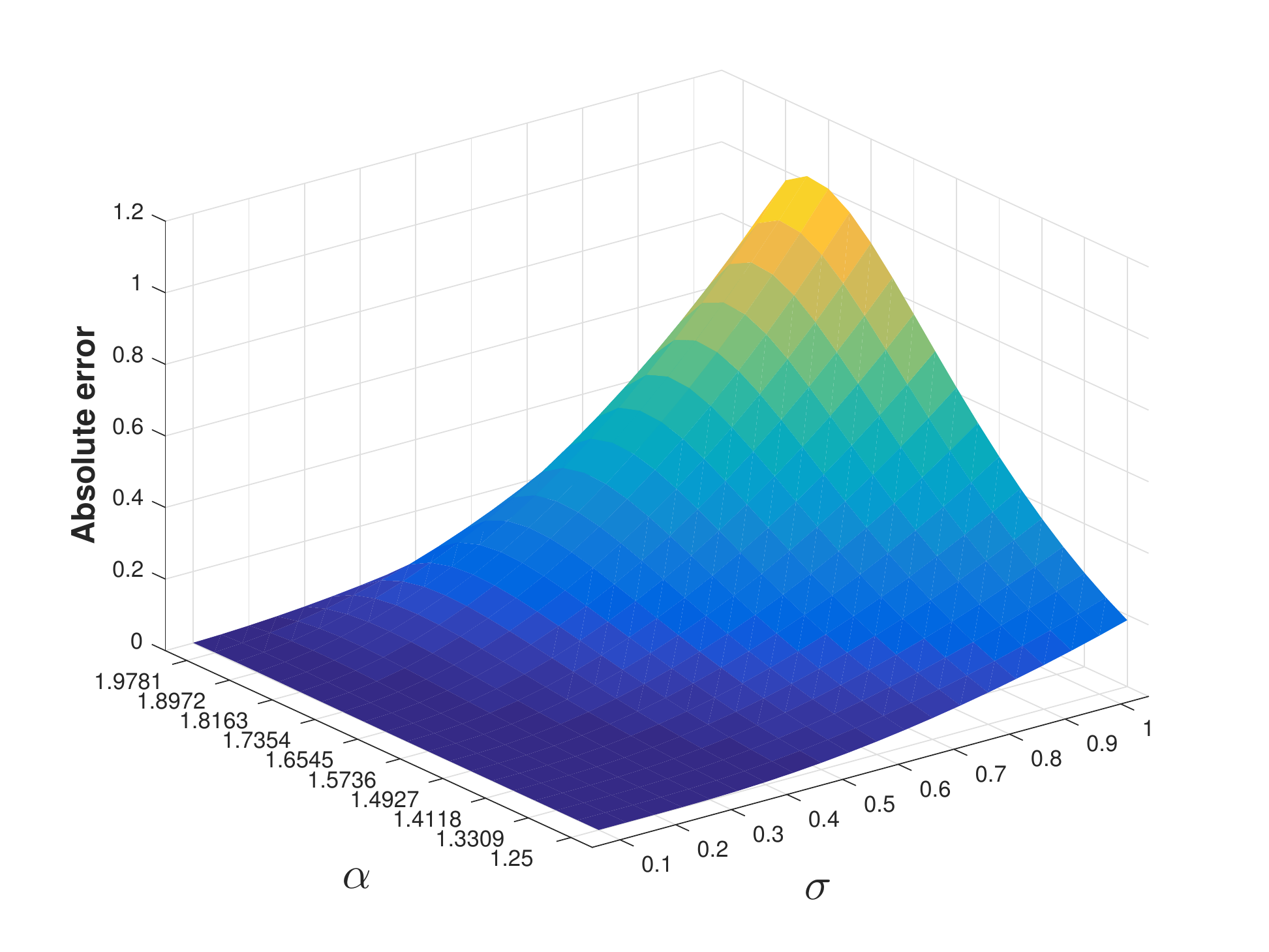}

}\subfloat[Relative error]{\includegraphics[width=0.5\textwidth]{6_Users_axel_Desktop_paper_cev_paraqlosubasaarx___os___concoverlett_error_abssoluto_surf_flat.pdf}

}

\caption{Absolute and relative error of of the path integral approach with
$T=0.5$, $r=0.05$, $S_{0}=100$ and $E=110$\label{fig:Absolute-and-relative-error-T05}}
\end{figure}

For a clearer and complete view, we present the continuous results
in Figs. \ref{fig:Comparisonsigma02T05}, \ref{fig:Comparisonsigma5T05}
and \ref{fig:Comparisonsigma9T05} The pricing and the running time
are showed for both models sweeping on values of $\alpha$. The figures
confirm the observed concussions in Table \ref{tab:T05} in the sense
that the running times of the proposed method of path integral are
significantly lower (right-hand side figures) than the traditional
solution methodology for the CEV model, especially when $\alpha$
tends to 2 where the time of the benchmark method rises considerably.
In terms of the accuracy, we can see that the path integral method
fits very well in all cases. In order to have an estimation of the
path integral approach, in the Fig. \ref{fig:Absolute-and-relative-error-T05}
the absolute and relative errors are shown for several values of $\alpha$
and $\sigma$. Always, the relative relative error is no longer that
10\% for the assumed parameters.

\begin{figure}[H]
\subfloat[Option Pricing]{\includegraphics[width=0.5\textwidth]{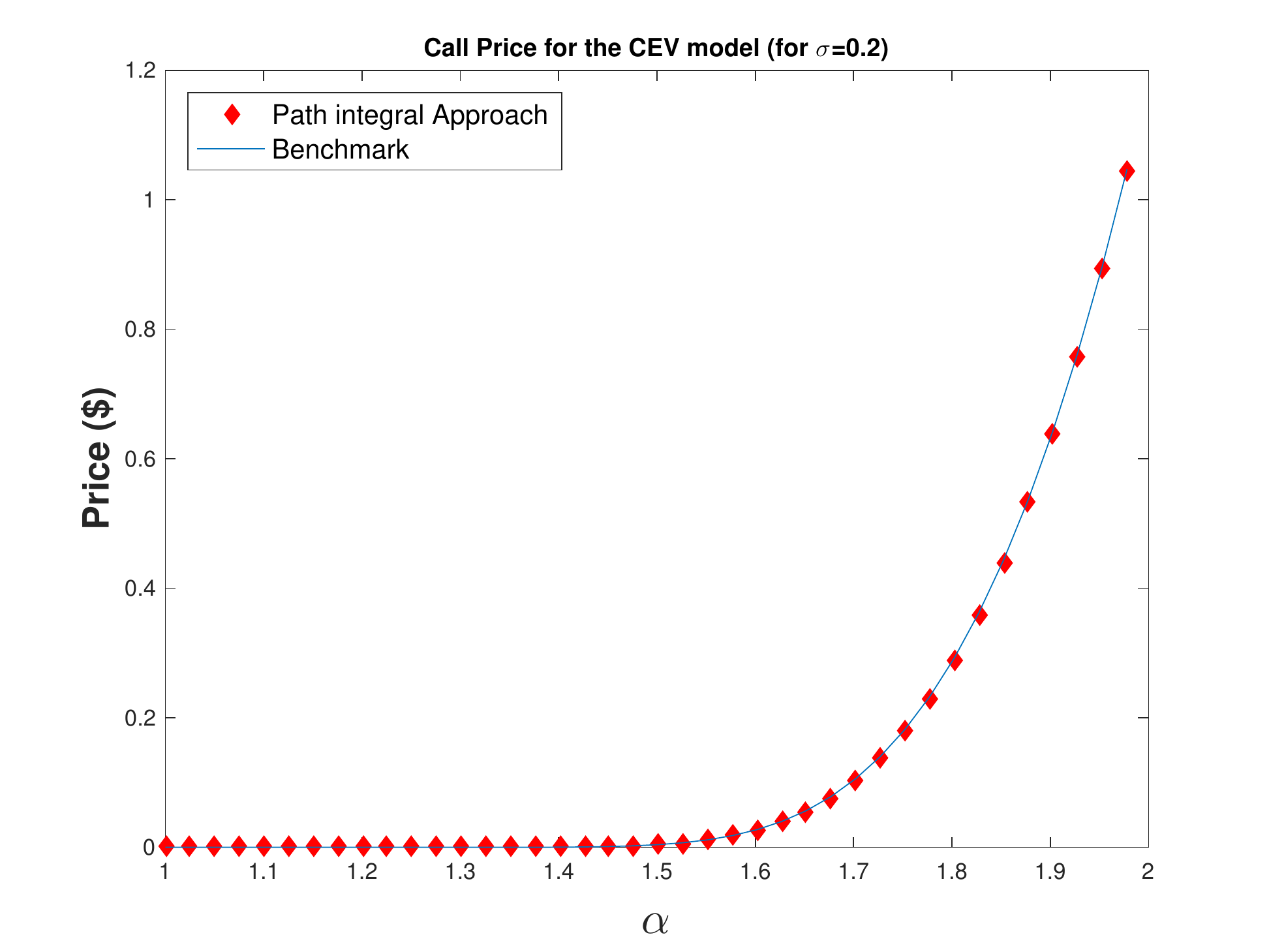}

}\subfloat[Running time]{\includegraphics[width=0.5\textwidth]{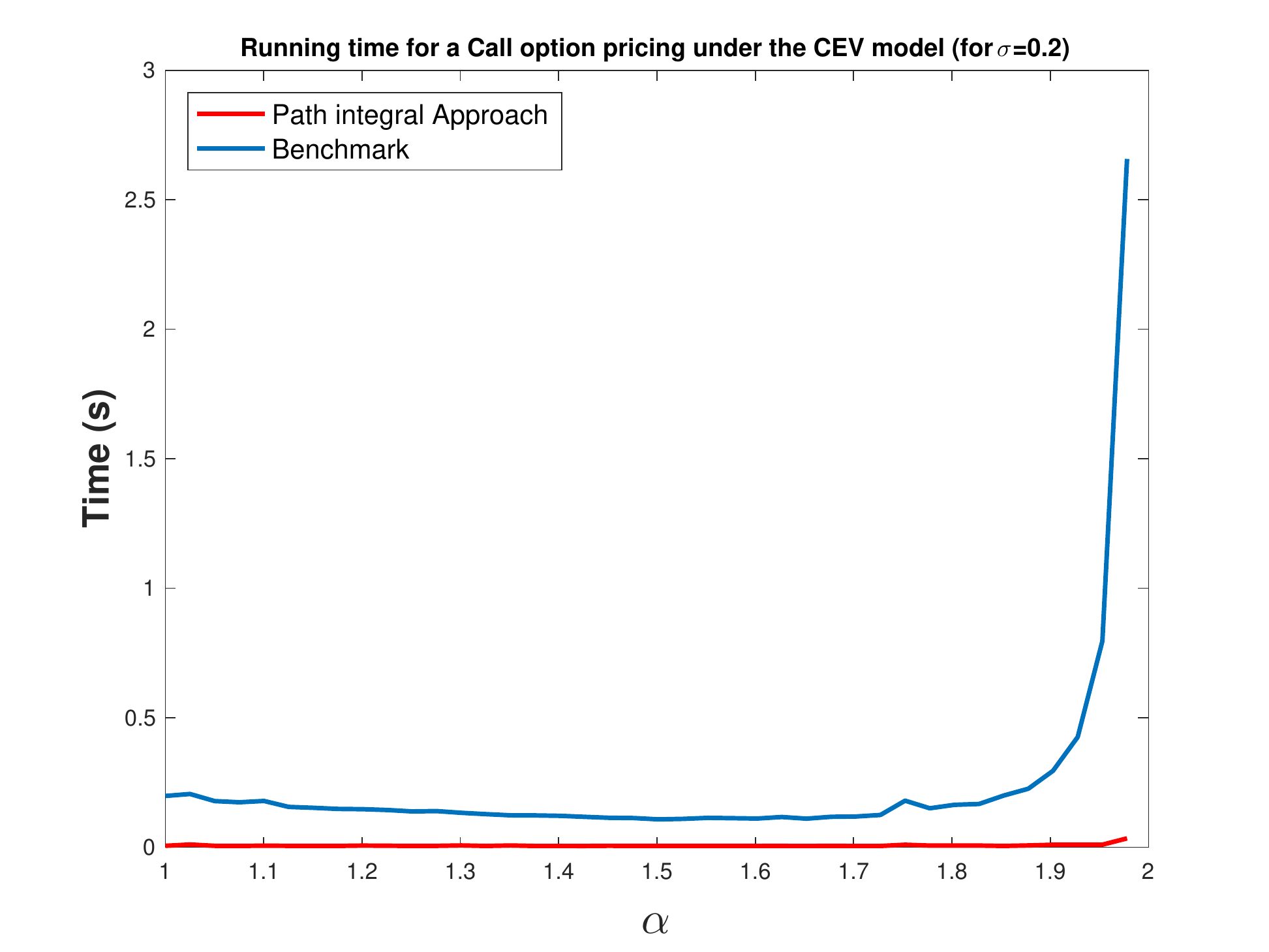}

}

\caption{Pricing and computational time for a Call option using $\sigma=$20\%,
$T=0.25$, $r=0.05$, $S_{0}=100$ and $E=110$ \label{fig:Comparisonsigma02T025}}
\end{figure}

\begin{figure}[H]
\subfloat[Option Pricing]{\includegraphics[width=0.5\textwidth]{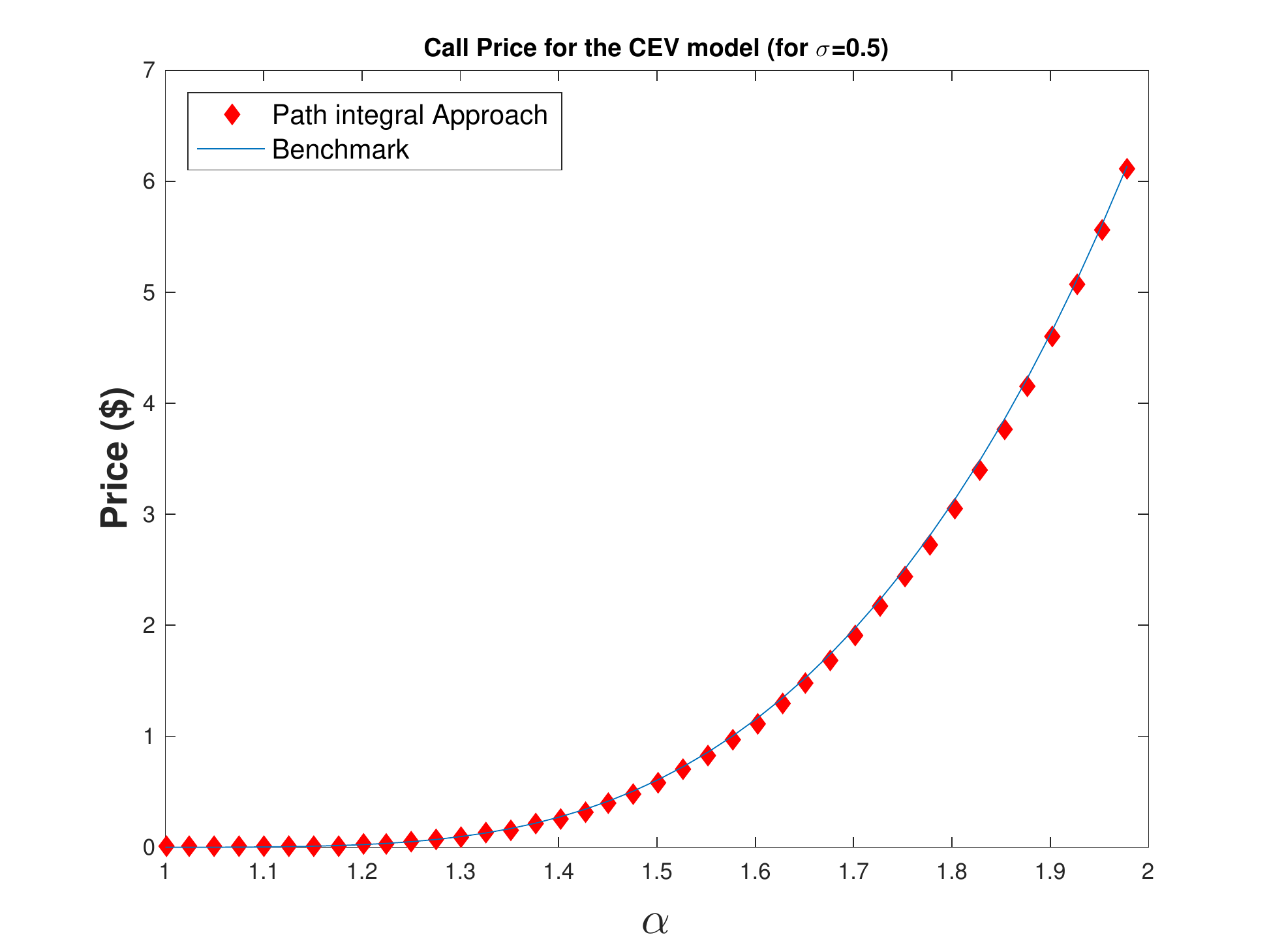}

}\subfloat[Running time]{\includegraphics[width=0.5\textwidth]{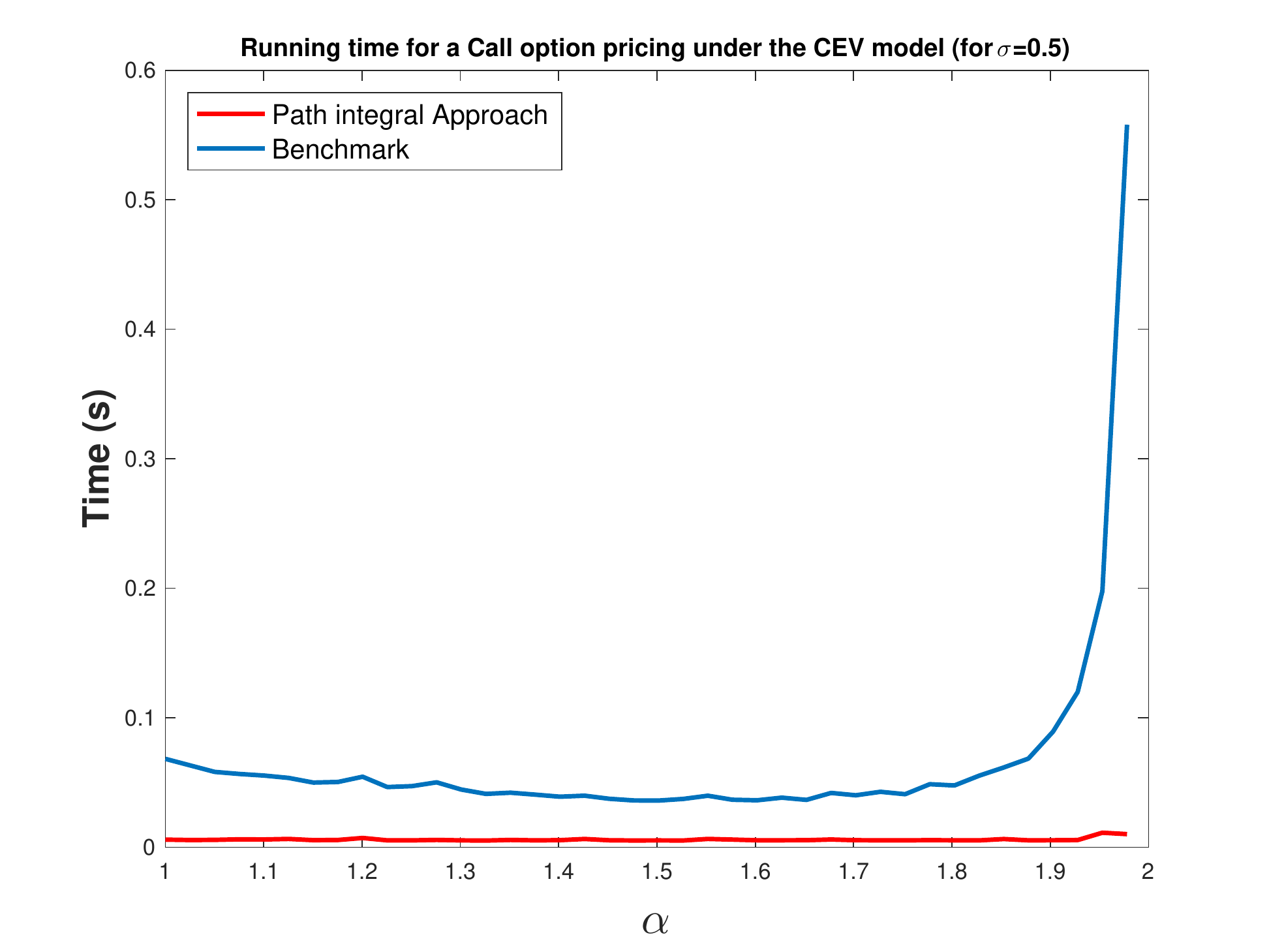}

}

\caption{Pricing and computational time for a Call option using $\sigma=$50\%,
$T=0.25$, $r=0.05$, $S_{0}=100$ and $E=110$\label{fig:Comparisonsigma5T025}}
\end{figure}

\begin{figure}[H]
\subfloat[Option Pricing]{\includegraphics[width=0.5\textwidth]{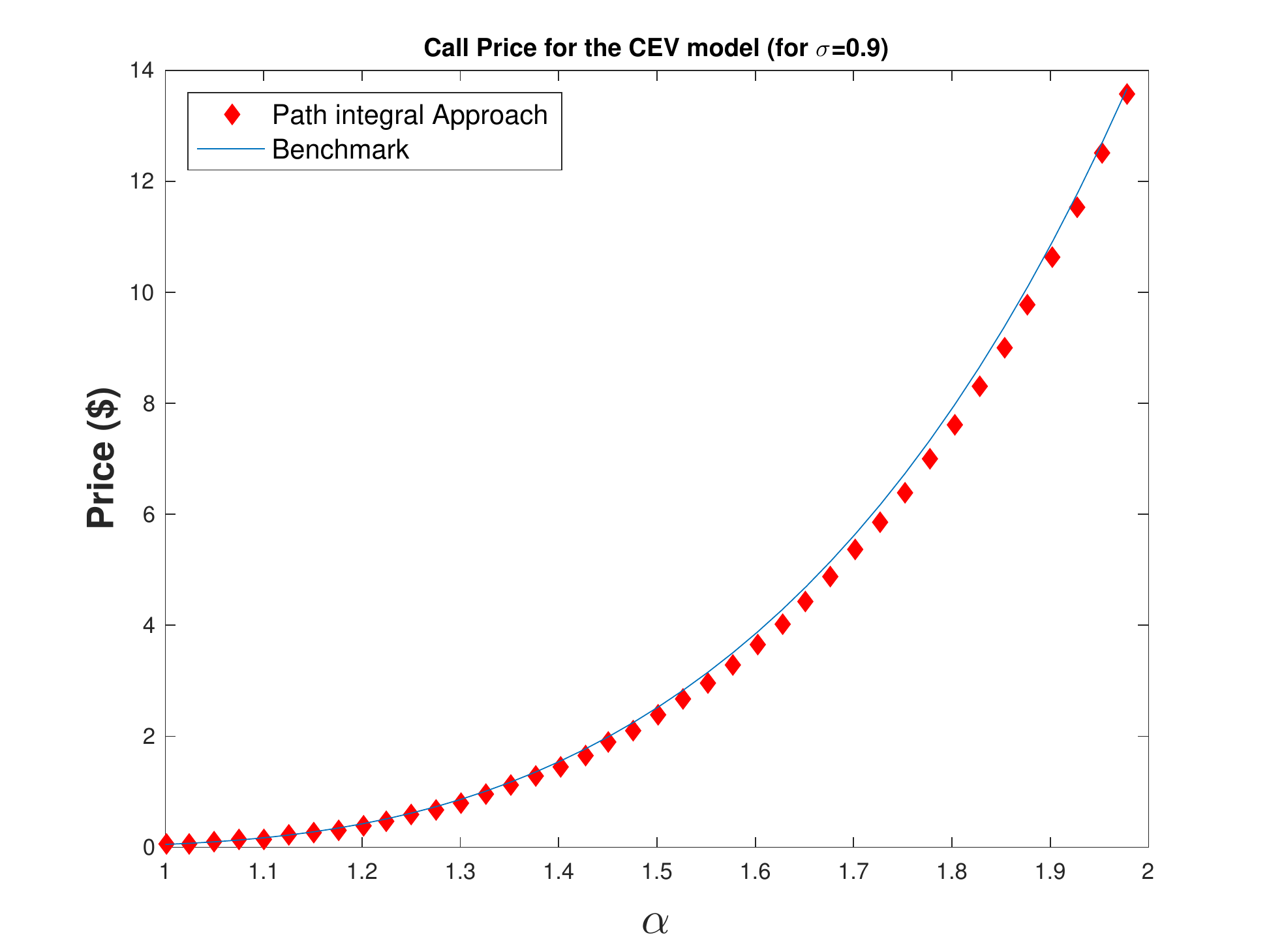}

}\subfloat[Running time]{\includegraphics[width=0.5\textwidth]{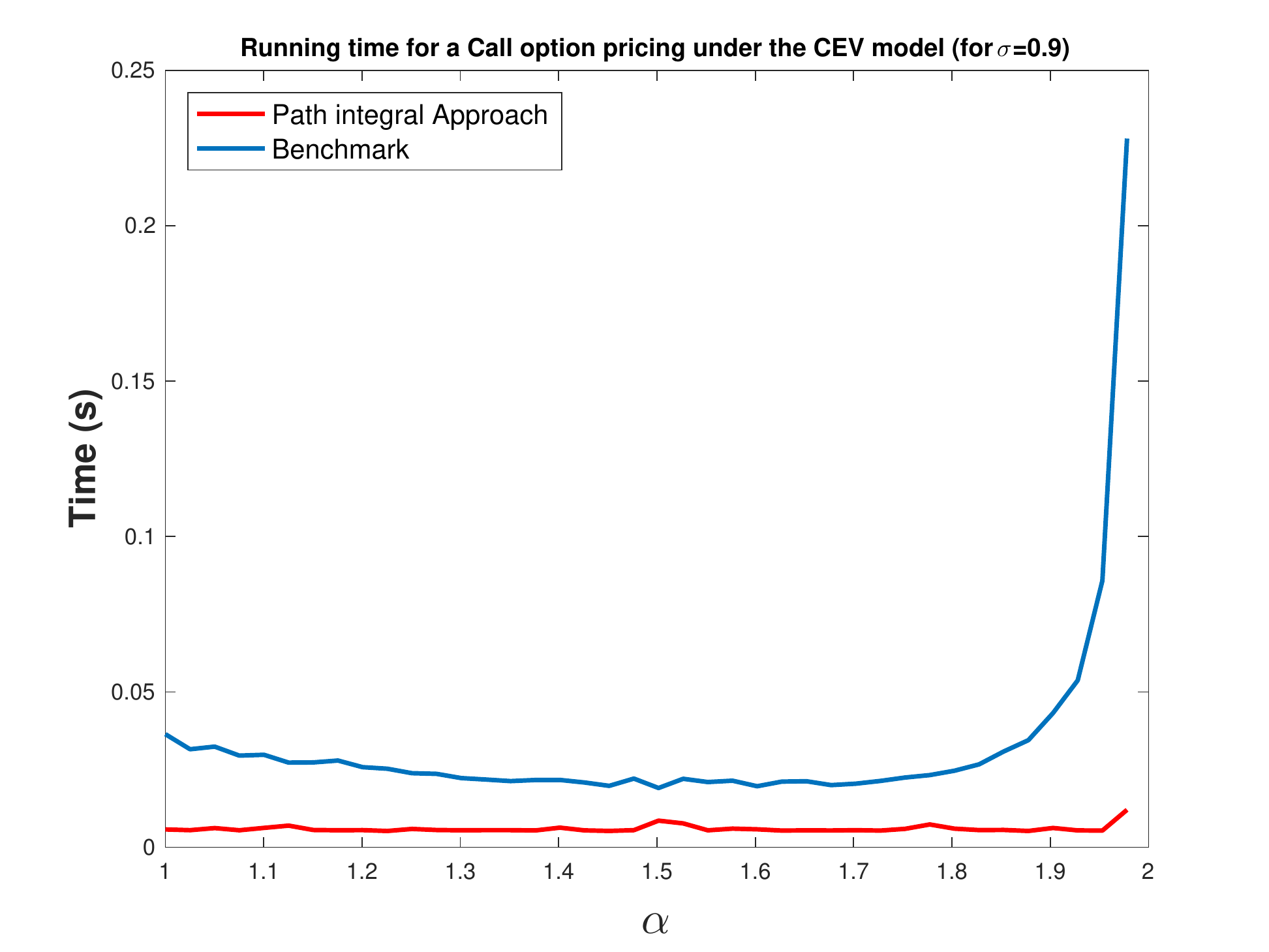}

}

\caption{Pricing and computational time for a Call option using $\sigma=$90\%,
$T=0.25$, $r=0.05$, $S_{0}=100$ and $E=110$ \label{fig:Comparisonsigma9T025}}
\end{figure}

\begin{figure}[H]
\subfloat[Absolute error]{\includegraphics[width=0.5\textwidth]{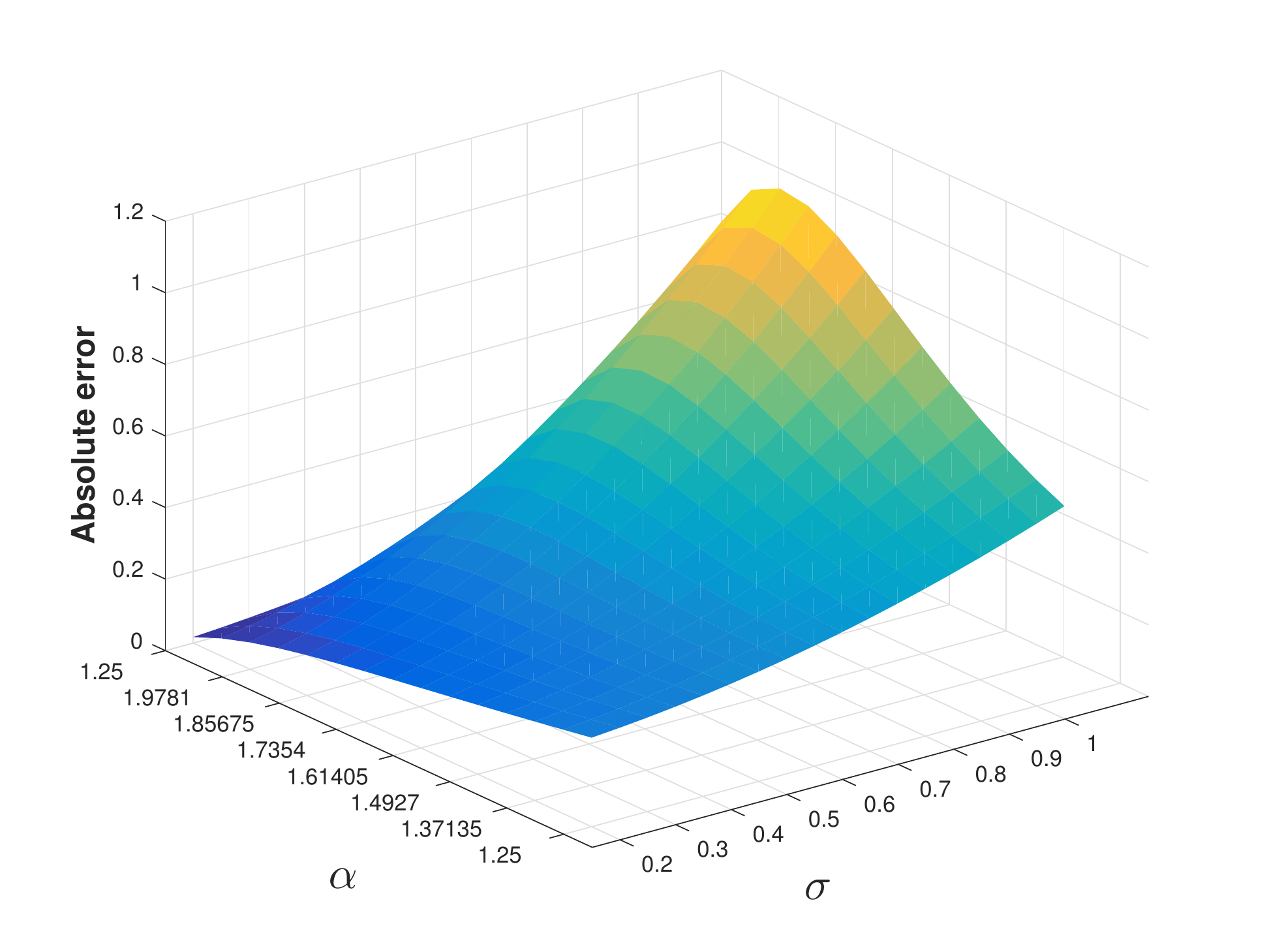}

}\subfloat[Relative error]{\includegraphics[width=0.5\textwidth]{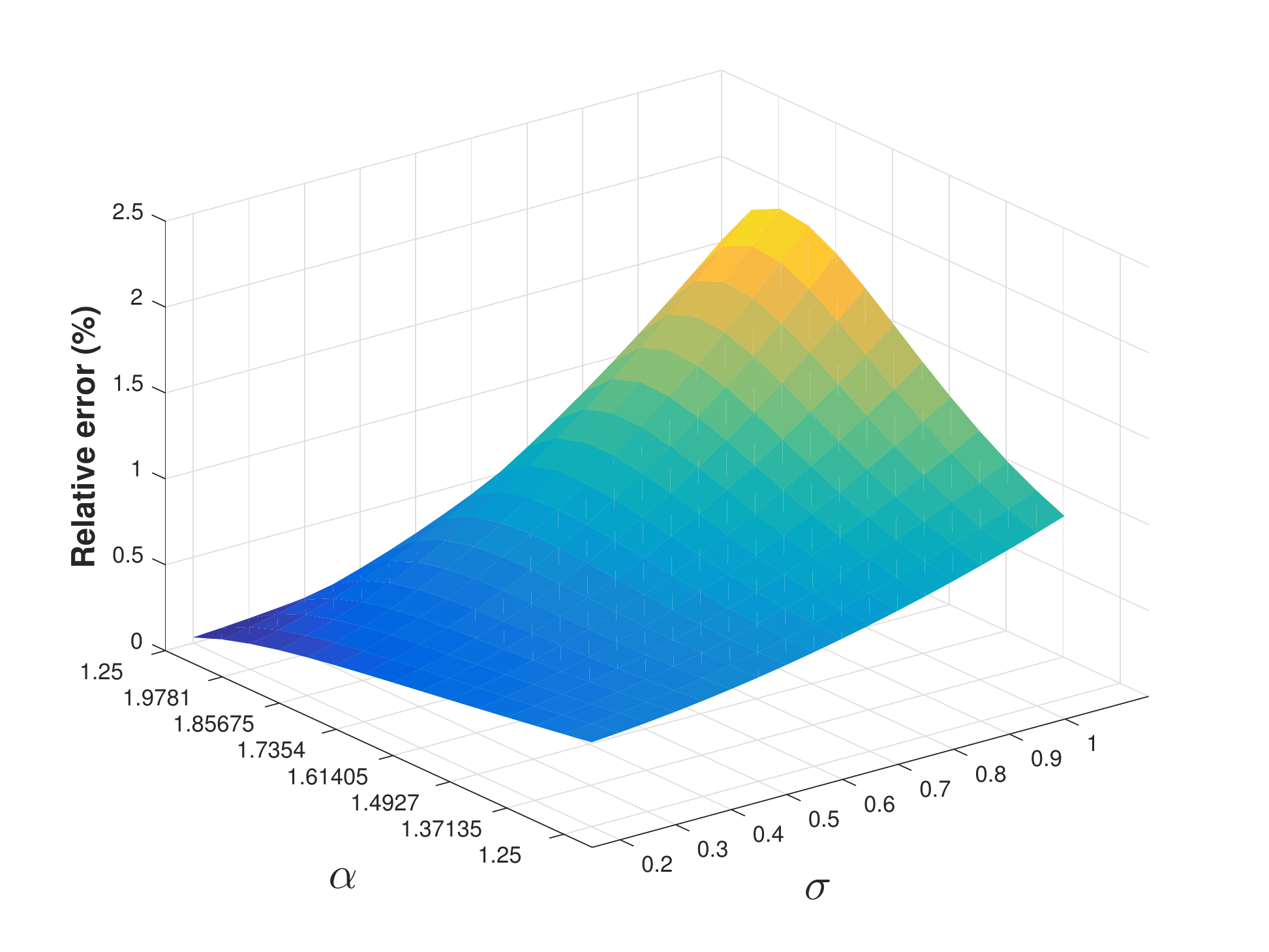}

}

\caption{Absolute and relative error of of the path integral approach with
$T=0.25$, $r=0.05$, $S_{0}=100$ and $E=110$\label{fig:Absolute-and-relative-error-T025}}
\end{figure}

If we use a lower time to maturity (three months) the results in terms
of computational time are very similar to the case $T=0.5$, and the
fit is still good too. In fact, for lower maturity the path integral
method performs better because the error is no greater than 2\%. This
is showed from Fig. \ref{fig:Comparisonsigma02T025} to Fig. \ref{fig:Absolute-and-relative-error-T025}.

\begin{figure}[H]
\subfloat[Option Pricing]{\includegraphics[width=0.5\textwidth]{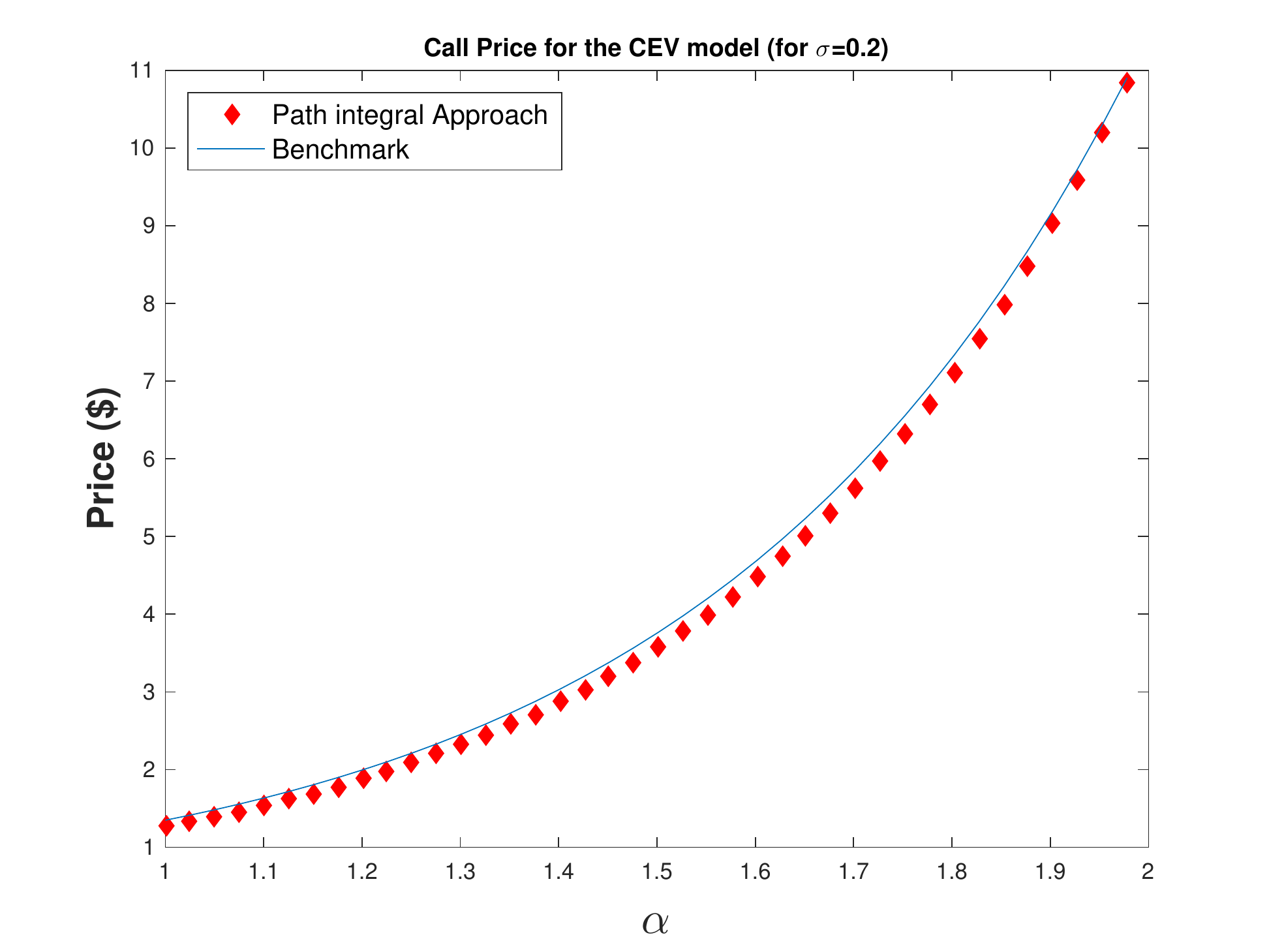}

}\subfloat[Running time]{\includegraphics[width=0.5\textwidth]{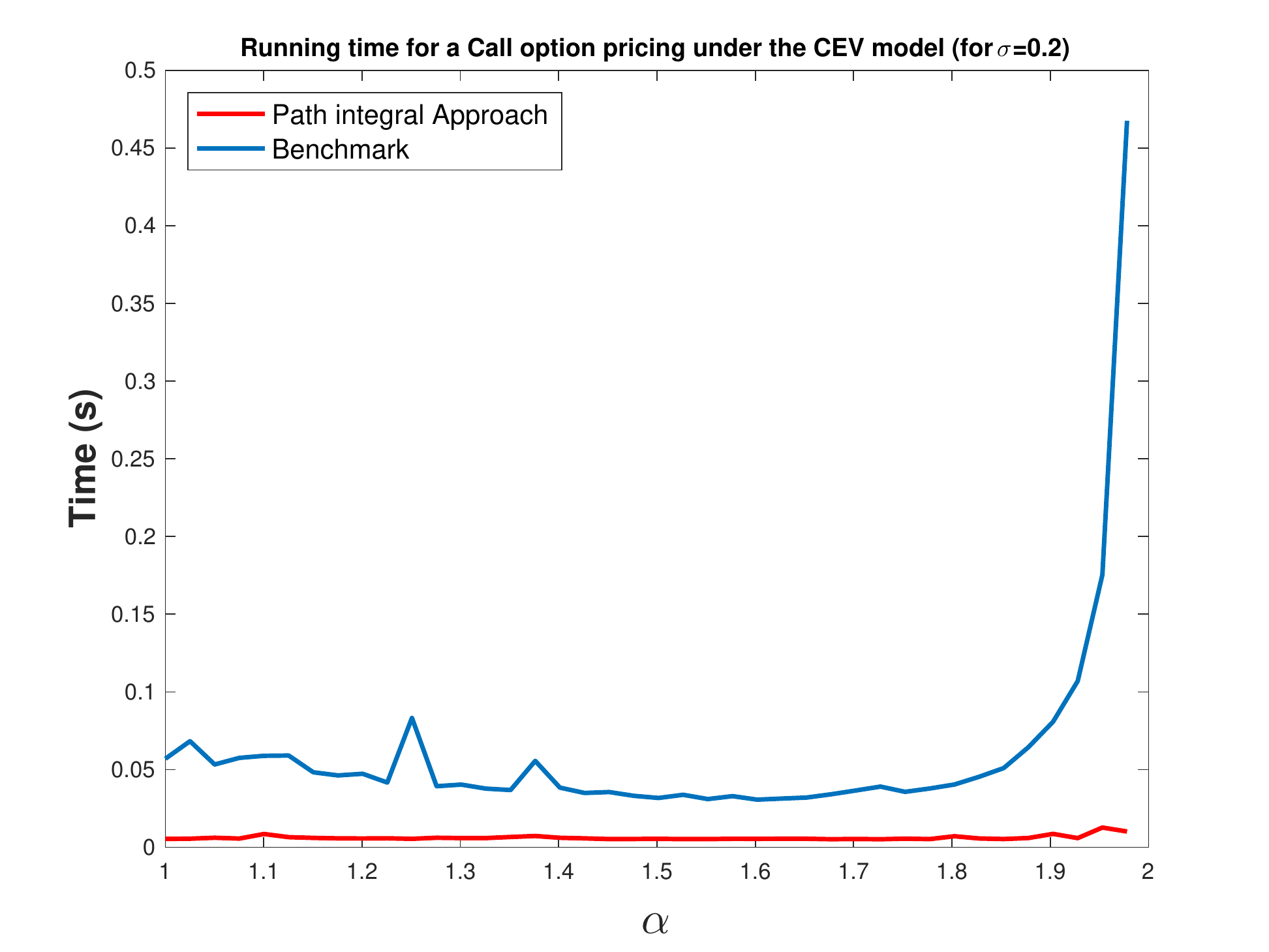}

}

\caption{Pricing and computational time for a Call option using $\sigma=$20\%,
$T=2$, $r=0.05$, $S_{0}=100$ and $E=110$ \label{fig:Comparisonsigma02T2}}
\end{figure}

\begin{figure}[H]
\subfloat[Option Pricing]{\includegraphics[width=0.5\textwidth]{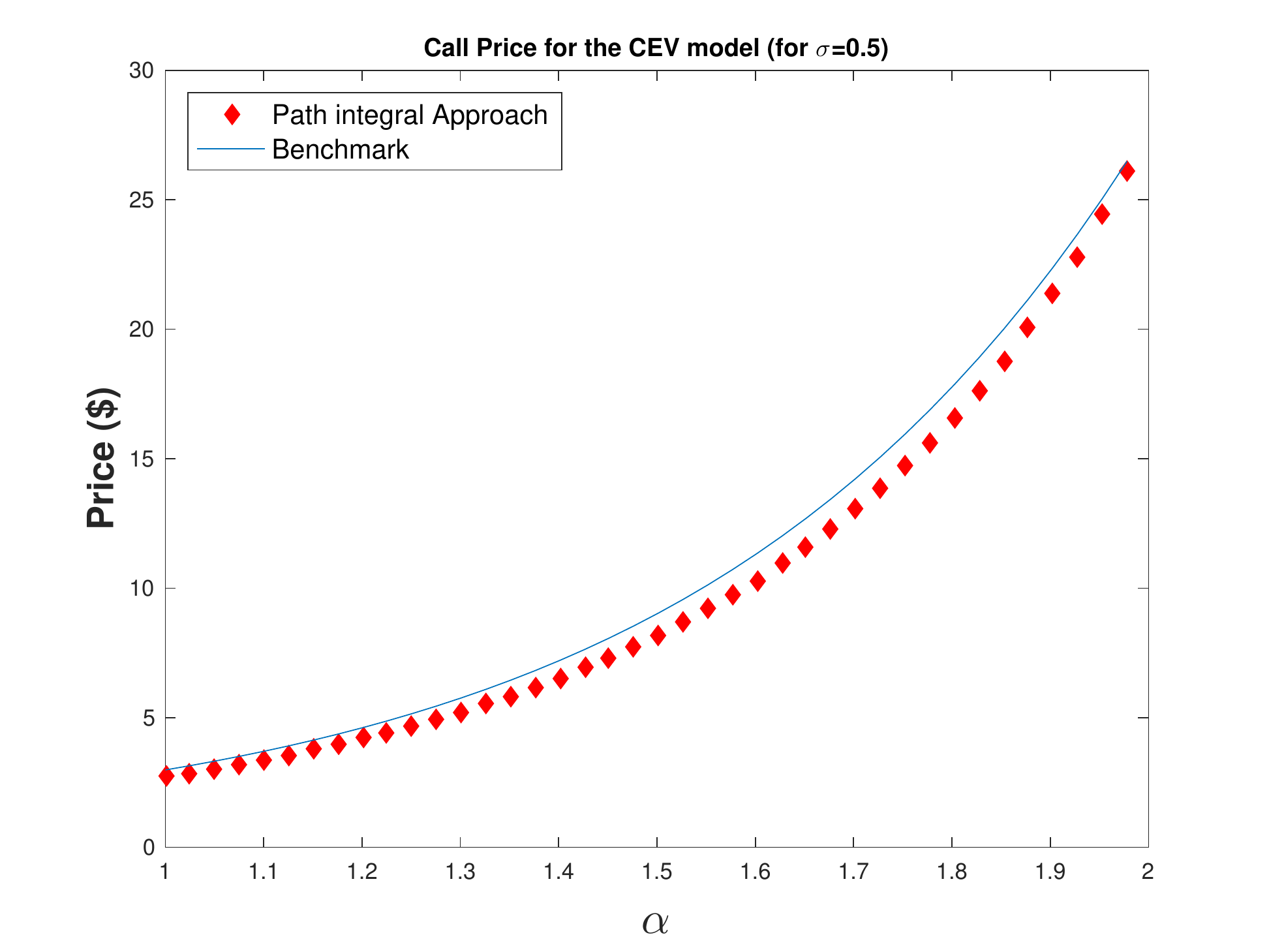}

}\subfloat[Running time]{\includegraphics[width=0.5\textwidth]{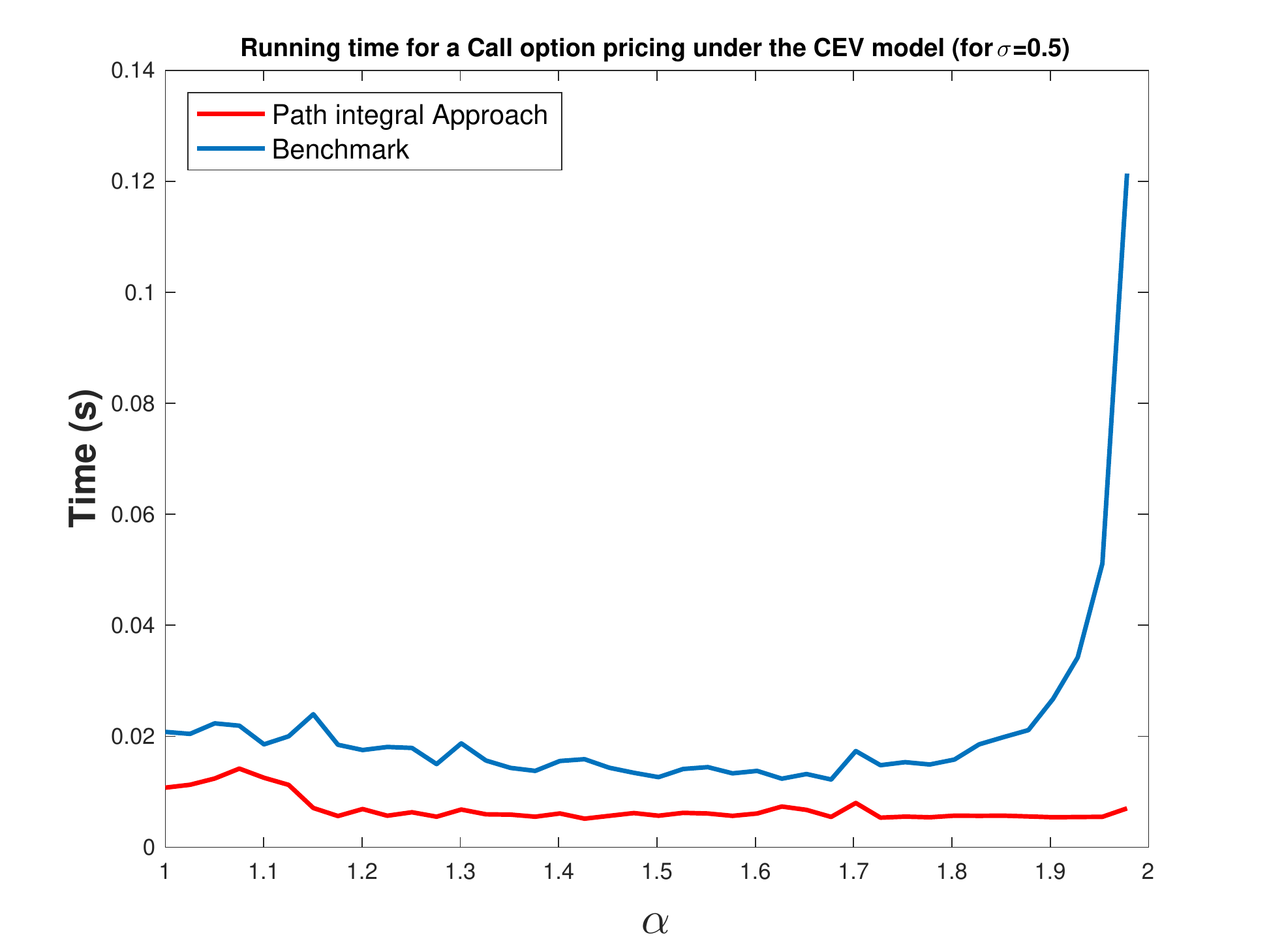}

}

\caption{Pricing and computational time for a Call option using $\sigma=$50\%,
$T=2$, $r=0.05$, $S_{0}=100$ and $E=110$\label{fig:Comparisonsigma5T2}}
\end{figure}

\begin{figure}[H]
\subfloat[Option Pricing]{\includegraphics[width=0.5\textwidth]{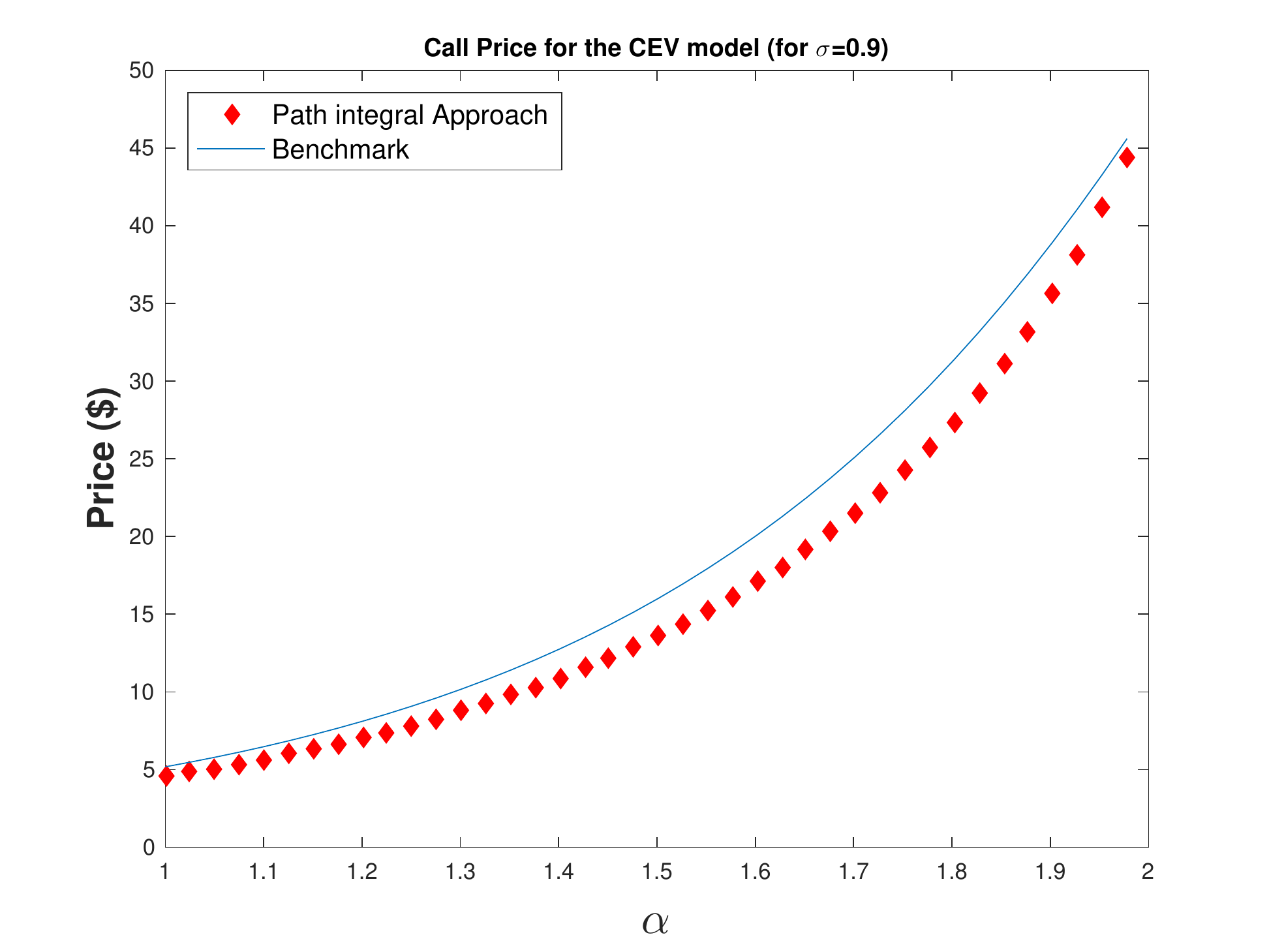}

}\subfloat[Running time]{\includegraphics[width=0.5\textwidth]{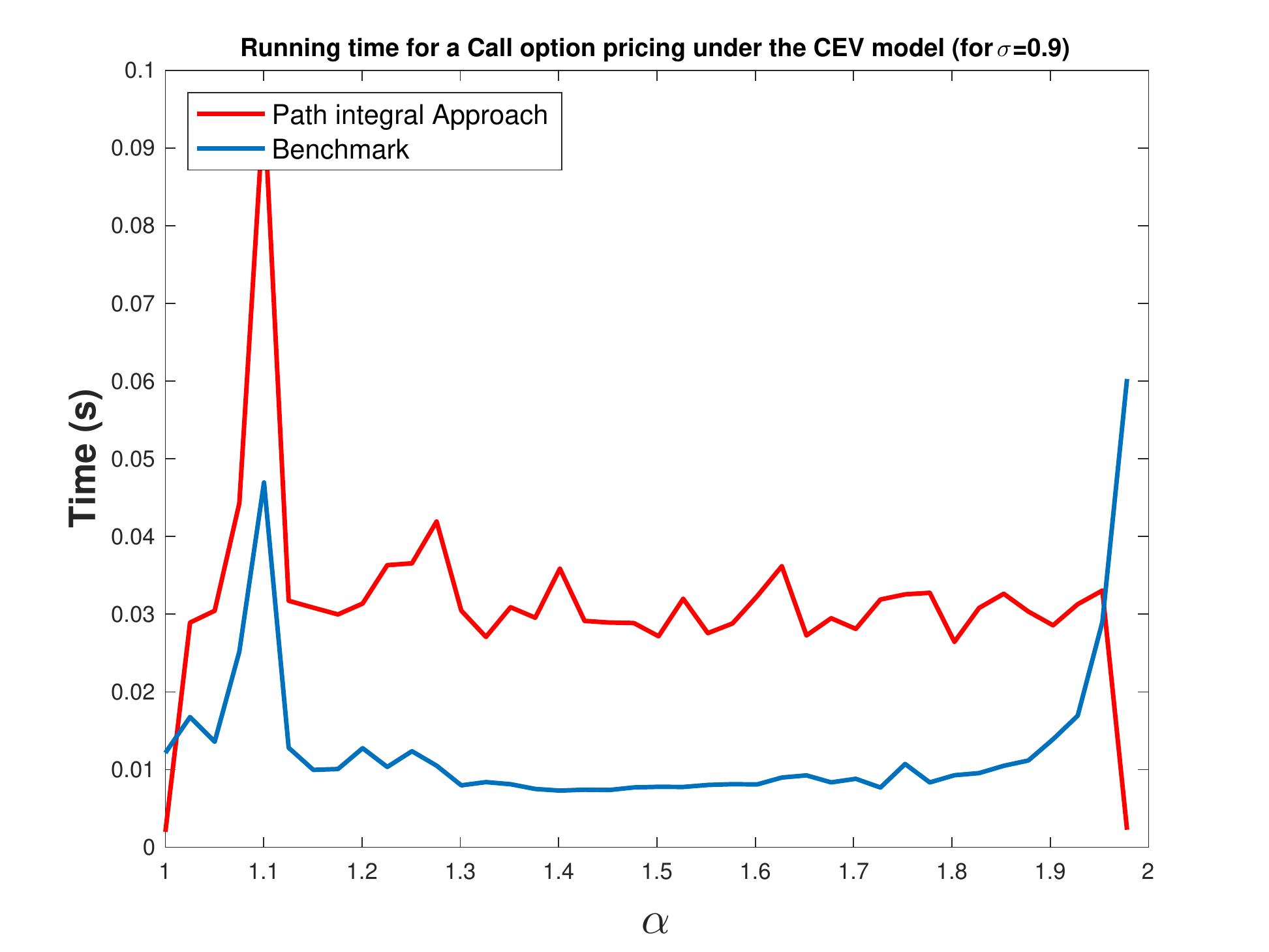}

}

\caption{Pricing and computational time for a Call option using $\sigma=$90\%,
$T=2$, $r=0.05$, $S_{0}=100$ and $E=110$ \label{fig:Comparisonsigma9T2}}
\end{figure}

\begin{figure}[H]
\subfloat[Absolute error]{\includegraphics[width=0.5\textwidth]{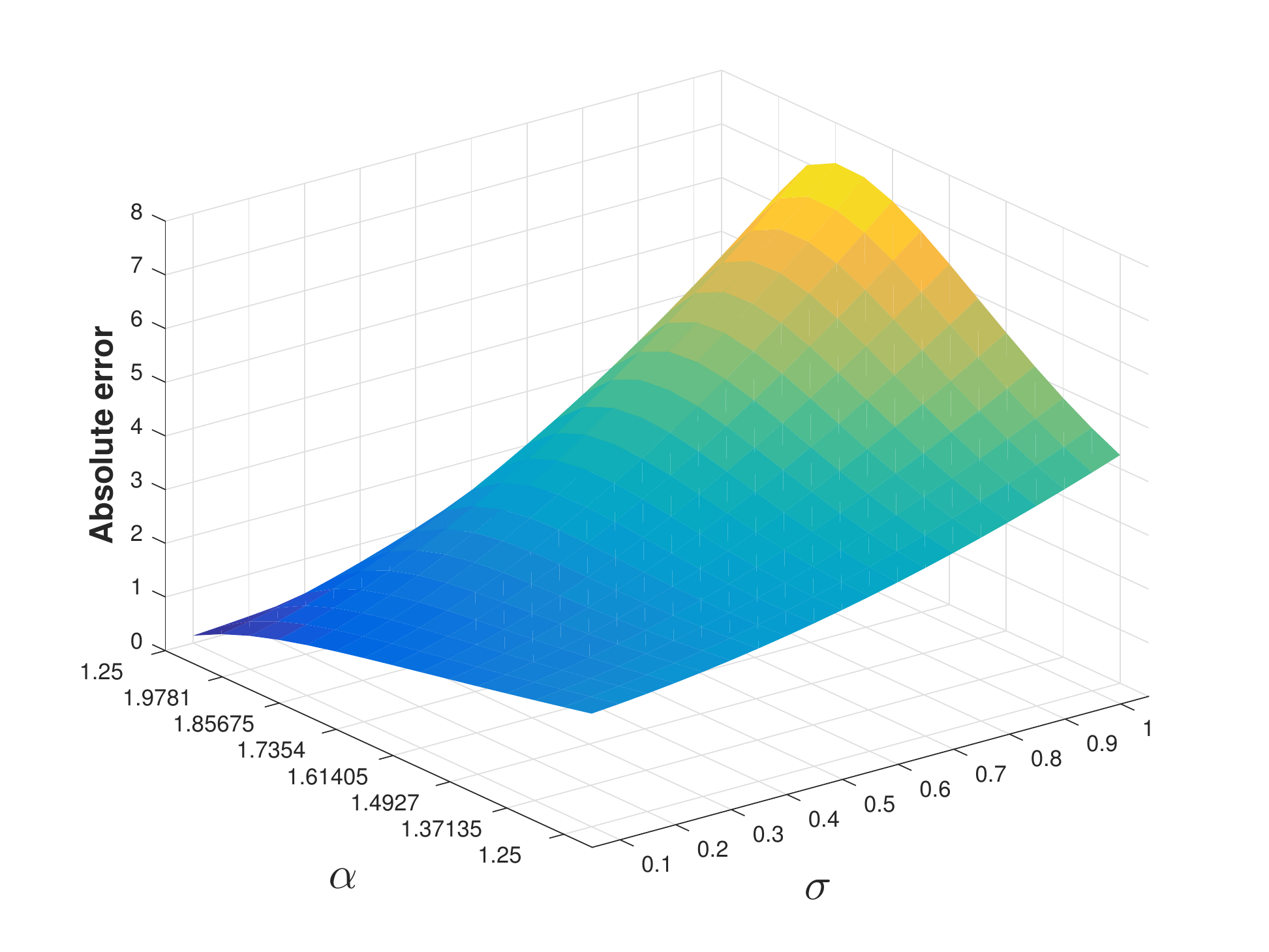}

}\subfloat[Relative error]{\includegraphics[width=0.5\textwidth]{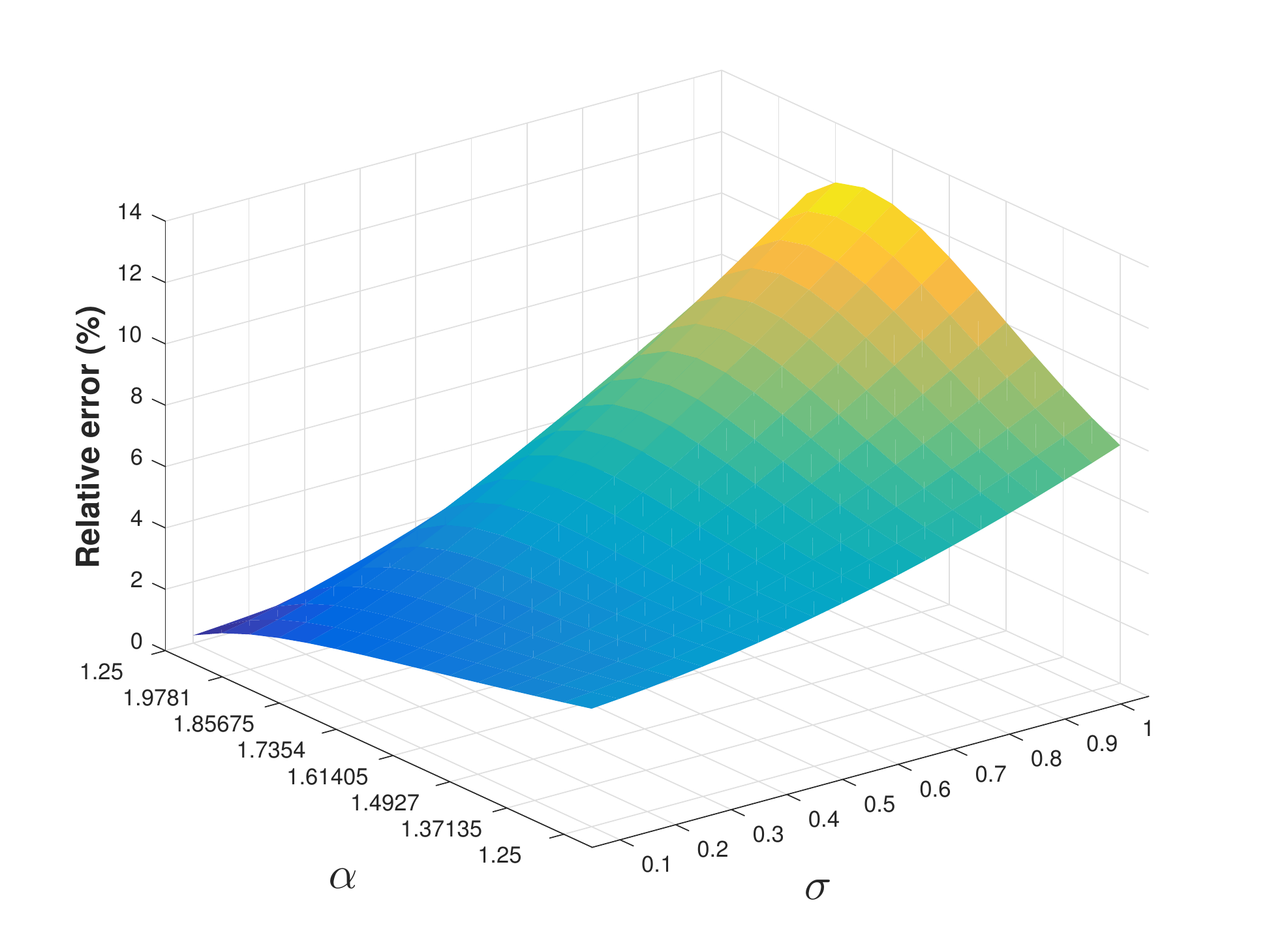}

}

\caption{Absolute and relative error of of the path integral approach with
$T=2$, $r=0.05$, $S_{0}=100$ and $E=110$\label{fig:Absolute-and-relative-error-T2}}
\end{figure}

For greater times to maturity, we have a change in the results, indicating
the limits of the semiclassical approximation. For a two years maturity
Figs. \ref{fig:Comparisonsigma02T2}-\ref{fig:Absolute-and-relative-error-T2}
we find results very similar to that of the previous cases, but with
a more deviation in the pricing (absolute error). Still, the relative
error remains lower than 12\%. However the figures show an increase
in the running time, being comparable the times of both approaches,
specially when the volatility rises and the elasticity is low. This
fact is confirmed when we use a maturity equals to 4 years. Indeed,
the computational cost increase, being the proposed method still competitive
for higher $\alpha$. In the same way, the pricing error goes up,
despite the fact that the relative error remains under 20\%. 

\begin{figure}[H]
\subfloat[Option Pricing]{\includegraphics[width=0.5\textwidth]{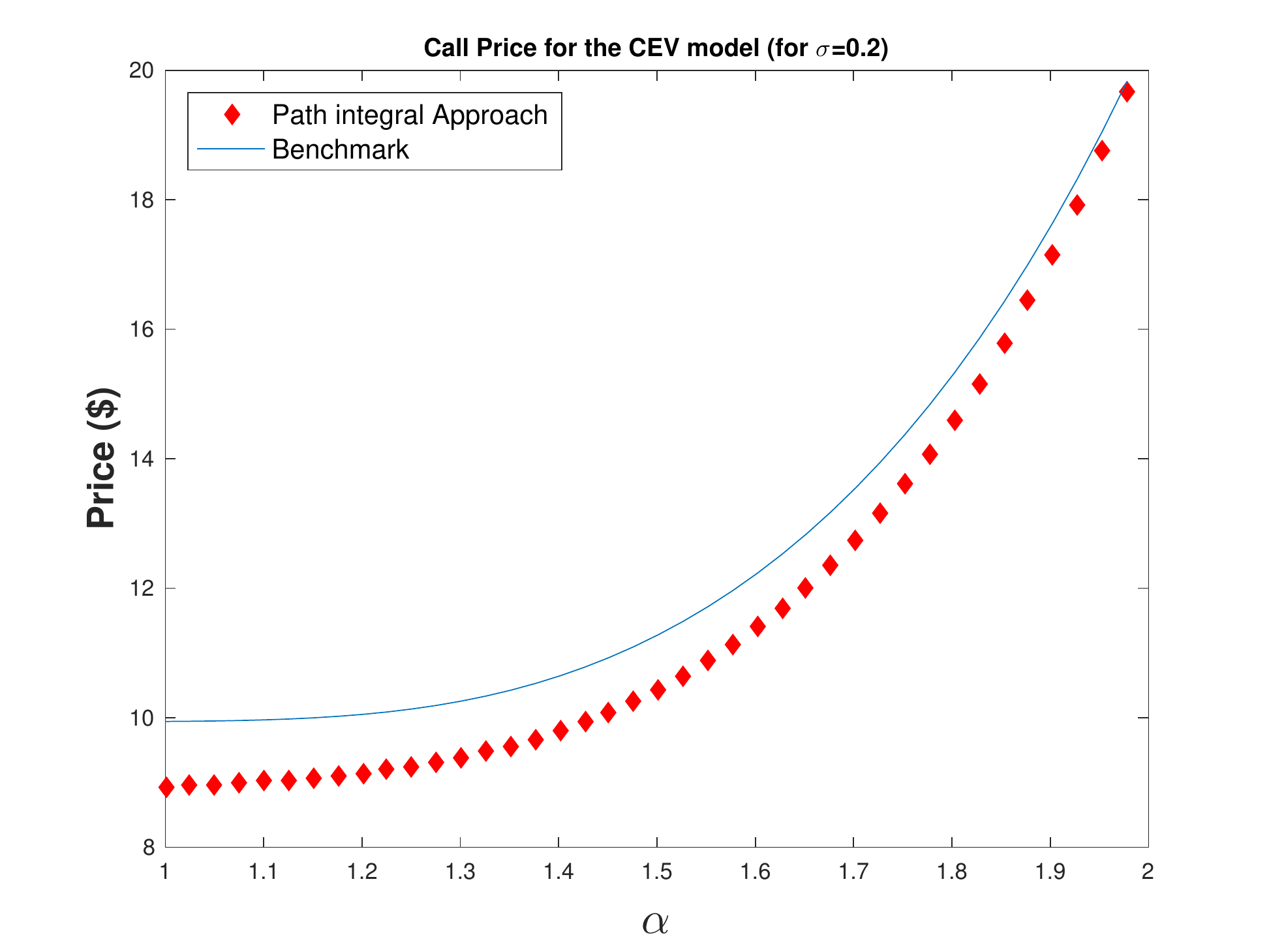}

}\subfloat[Running time]{\includegraphics[width=0.5\textwidth]{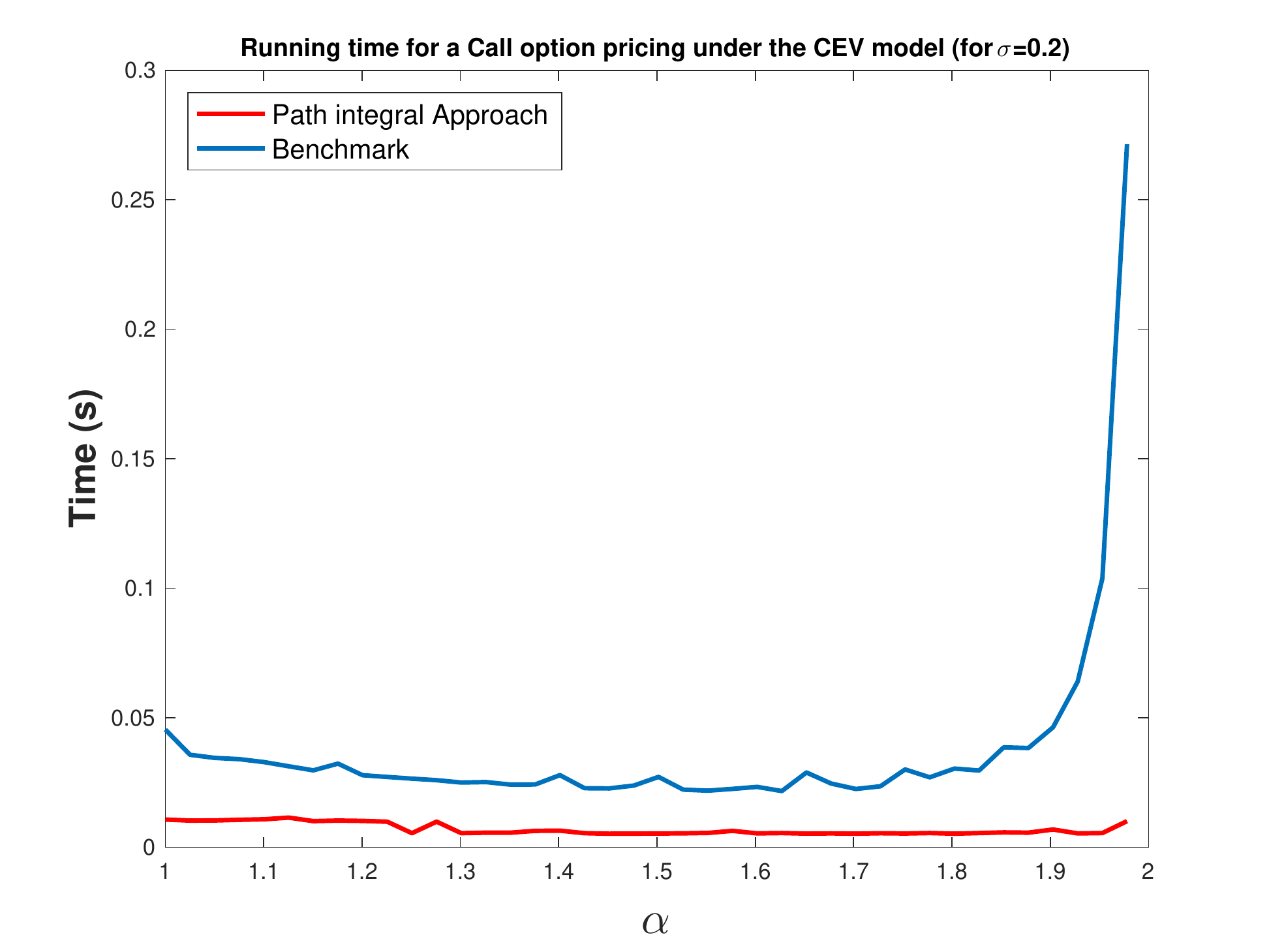}

}

\caption{Pricing and computational time for a Call option using $\sigma=$20\%,
$T=4$, $r=0.05$, $S_{0}=100$ and $E=110$ \label{fig:Comparisonsigma02T2-1}}
\end{figure}

\begin{figure}[H]
\subfloat[Option Pricing]{\includegraphics[width=0.5\textwidth]{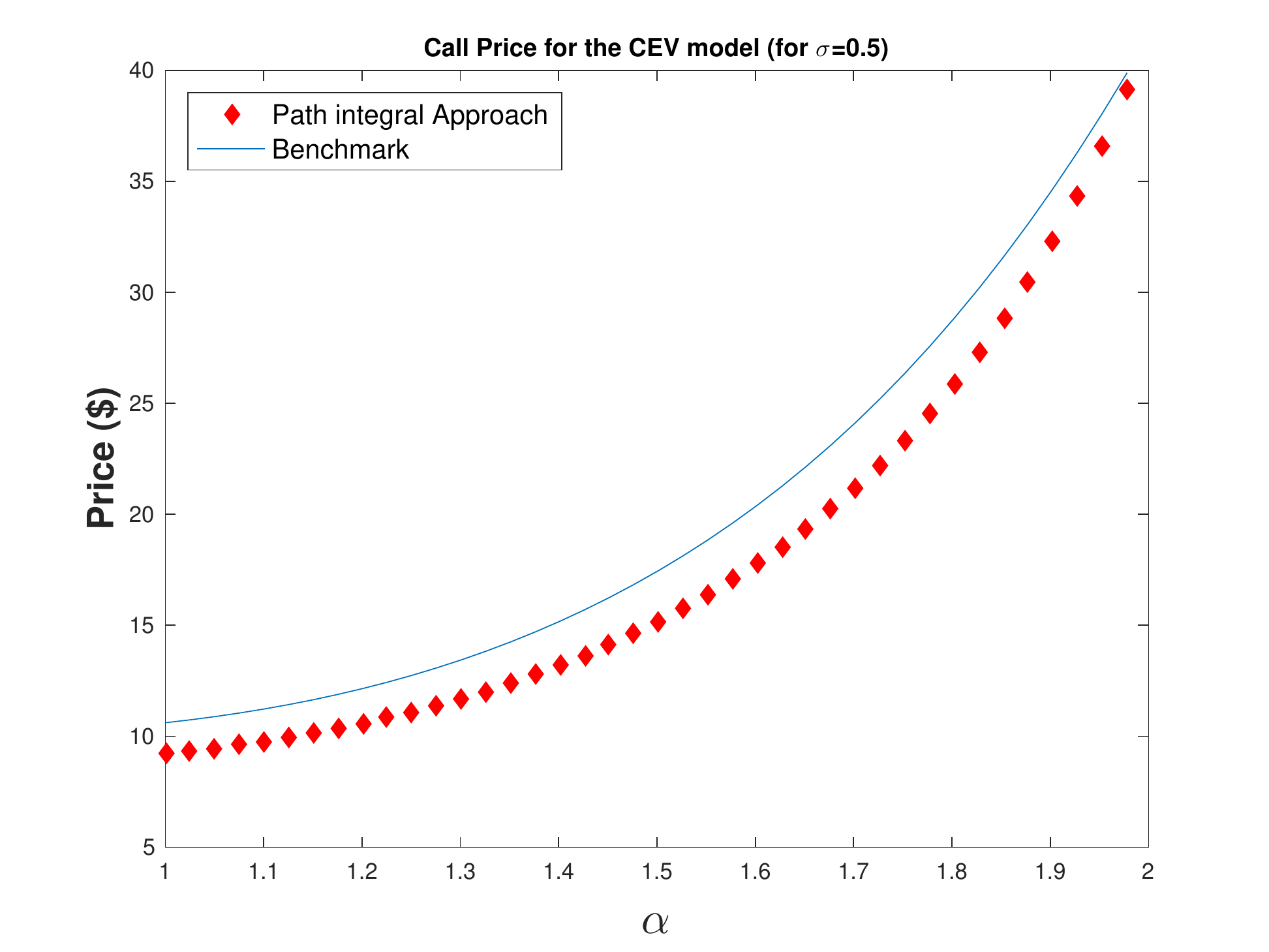}

}\subfloat[Running time]{\includegraphics[width=0.5\textwidth]{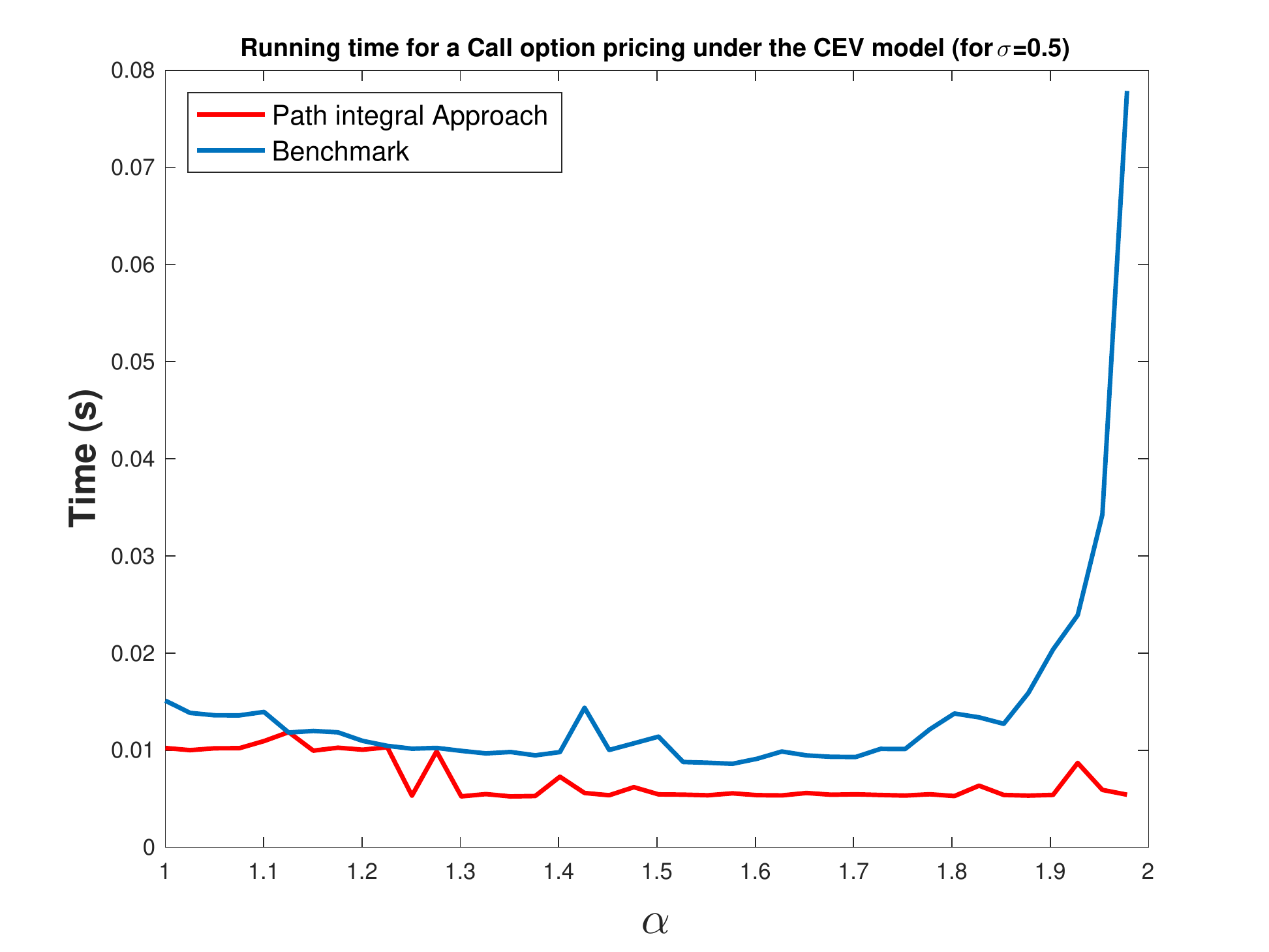}

}

\caption{Pricing and computational time for a Call option using $\sigma=$50\%,
$T=4$, $r=0.05$, $S_{0}=100$ and $E=110$\label{fig:Comparisonsigma5T2-1}}
\end{figure}

\begin{figure}[H]
\subfloat[Option Pricing]{\includegraphics[width=0.5\textwidth]{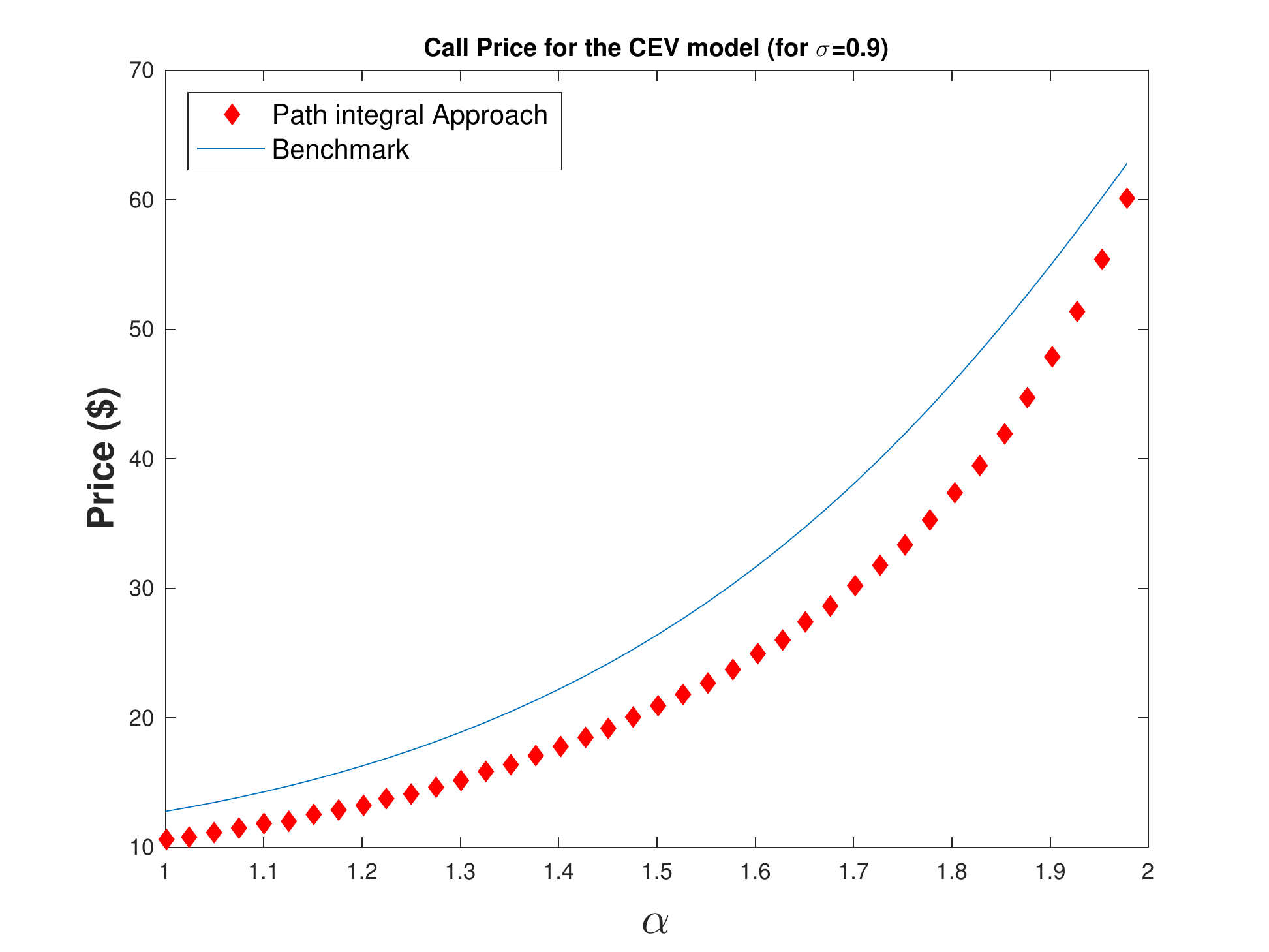}

}\subfloat[Running time]{\includegraphics[width=0.5\textwidth]{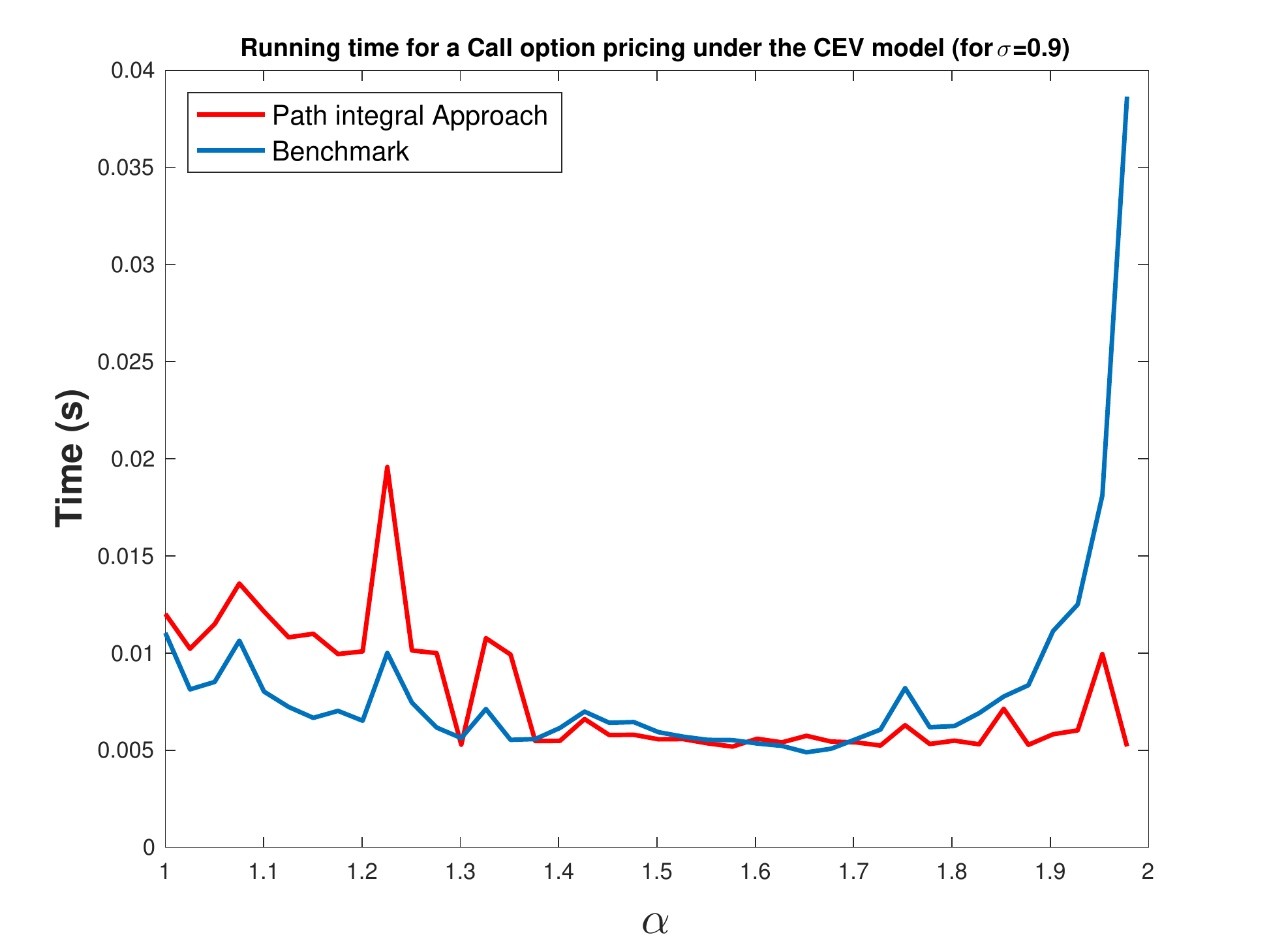}

}

\caption{Pricing and computational time for a Call option using $\sigma=$90\%,
$T=4$, $r=0.05$, $S_{0}=100$ and $E=110$ \label{fig:Comparisonsigma9T2-1}}
\end{figure}

\begin{figure}[H]
\subfloat[Absolute error]{\includegraphics[width=0.5\textwidth]{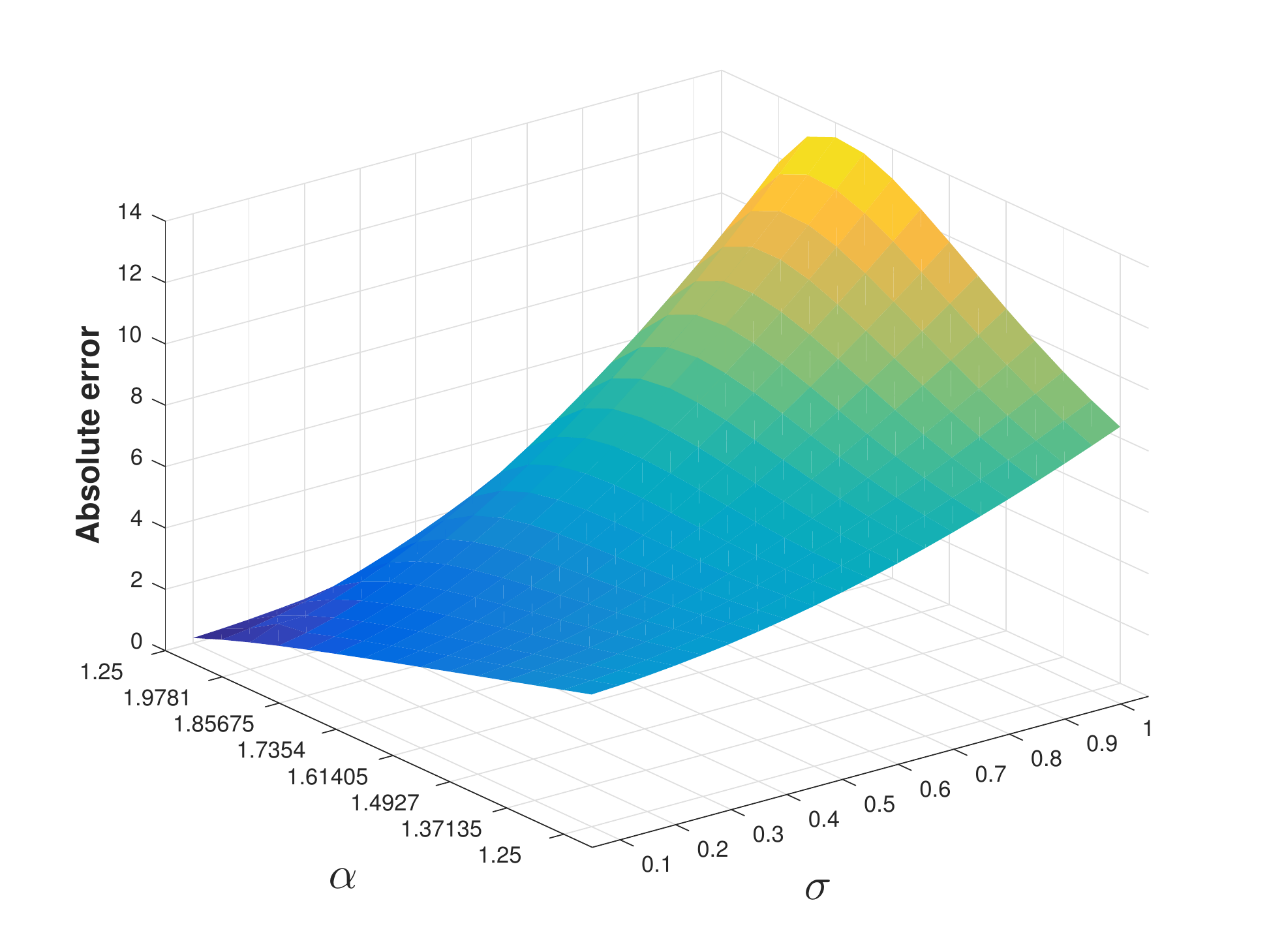}

}\subfloat[Relative error]{\includegraphics[width=0.5\textwidth]{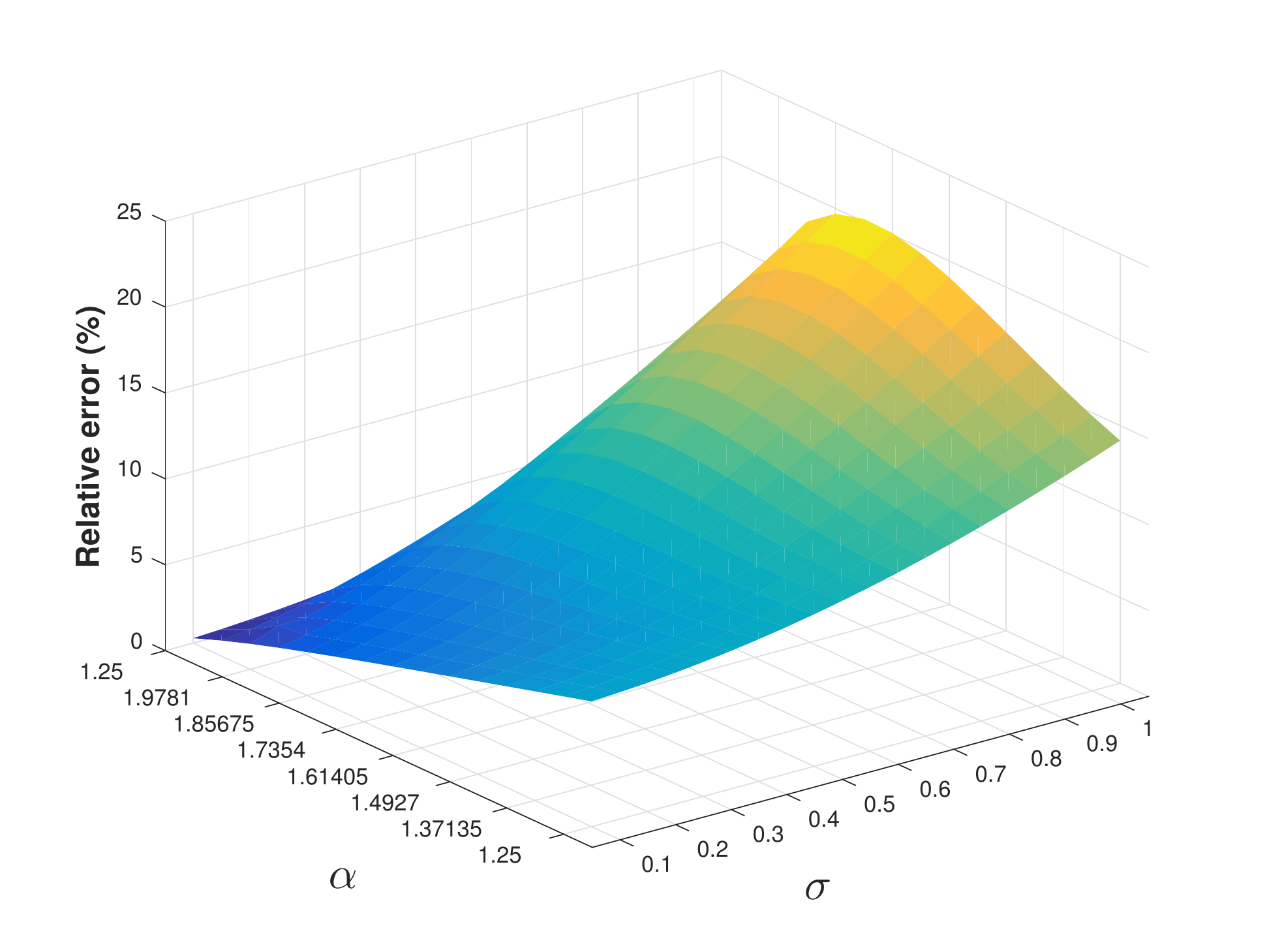}

}

\caption{Absolute and relative error of the path integral approach with $T=4$,
$r=0.05$, $S_{0}=100$ and $E=110$\label{fig:Absolute-and-relative-error-T2-1}}
\end{figure}

\section{Summary and further research}

In this paper, a new numerical method for computing the CEV model
was developed. In particular, this new approach was based on the semiclassical
approximation of Feynman's path integral model. This formulation dealt
with some of the limitations of the conventional approach based on
the non-central chi-squared distribution. 

The experimental results showed a good fit between the new proposed
method and the traditional methodology (setting the former as benchmark),
and also a lower computational cost, measured as the running time
of each model.

We analyze several hypothetical scenarios, using different maturities,
volatilities and elasticities. In most cases, the running time is
one order of magnitude lower than the benchmark, but if the elasticity
tends to one, this difference is higher. As an accuracy measure the
absolute and relative error are computed. For the range $10\%<\sigma<100\%$
and $1.25<\alpha<1.97$ the relative error is below than 20\% in all
the cases. Nevertheless, for short maturities and lower volatilities,
the error decreases considerably, coming to be less than 10\% for
small maturities (under 2\% for T=0.25!) and for $\sigma<50
$.

The main remark is that this novel methodology allow to evaluate an
European contract under the CEV model computing only an integral without
any complex numerically method. The accuracy and efficiency of this
method, positions it as a great competitor for the conventional method
based on the non-central chi-squared distribution. 

In terms of future research, a natural first extension of the paper
is to adapt the proposed methodology to American options. Also, the
pricing of exotic options would be a good target. Another interesting
research line is to apply the semiclassical approximation of Feynman's
path integral model to more sophisticated stochastic volatility models
such as: Heston, SABR or GARCH type models, where the traditional
current solutions are much more complicated than that of the CEV model,
and hence the potential value added of this methodology could be greater.\\

\bibliographystyle{unsrt}
\bibliography{63_Users_axel_Desktop_paper_cev_paraqlosubasaarxivyparatirarchivos___concoverlett_tesis}

\end{document}